\documentclass[%
 reprint,
 amsmath,amssymb,
 aps,longbibliography
]{revtex4-1}

\usepackage{graphicx}
\usepackage{dcolumn}
\usepackage{bm}
\usepackage{textcomp}
\usepackage{amsmath,amssymb,amsfonts}
\usepackage{graphicx}
\usepackage[percent]{overpic}
\usepackage{lipsum}
\usepackage{enumerate}
\usepackage{mathrsfs}
\usepackage{balance}
\usepackage{mathtools}
\usepackage{xr}
\usepackage{subfigure}
\usepackage{afterpage}
\usepackage{cases}
\usepackage{scalerel}
\usepackage{xcolor}
\usepackage{nicefrac,xfrac}

\newcommand\mydots{\ifmmode\ldots\else\makebox[1em][c]{.\hfil.\hfil.}\fi}

\begin{document}

\preprint{APS/123-QED}

\title{Approach to the Analysis and Synthesis of Cylindrical Metasurfaces
    with Non-circular Cross Sections Based on Conformal Transformations}

\author{Gengyu Xu}
 \email{paul.xu@mail.utoronto.ca}
\author{George V. Eleftheriades}%
\author{Sean V. Hum}%
\affiliation{%
The Edward S. Rogers Sr. Department of Electrical and Computer Engineering, University of Toronto,\\ Toronto, Ontario M5S 3H7, Canada}%

\date{\today}

\begin{abstract}
We present methods for analyzing and designing cylindrical electromagnetic metasurfaces with non-circular cross sections based on conformal transformations. It can be difficult to treat surfaces with non-canonical geometries since they generally do not admit straightforward solutions to the Helmholtz wave equation subject to the appropriate boundary conditions. This leads to the reliance on full wave numerical techniques which are only suitable for the analysis, but not the synthesis, of these surfaces. We address this issue by employing conformal transformations to map the physical space into a computational space in which the surface coincides with a circular cylinder. The electromagnetic boundary conditions on the surface remain intact under the transformations due to their angle-preserving nature. However, they are much more easily enforced. As a result, analytical modal solutions for the scattered fields are readily obtainable, which facilitate closed-form analysis and synthesis equations for general non-circular cylindrical metasurfaces. One important utility enabled by the proposed framework is the efficient identification of electromagnetic field distributions that satisfy local power conservation. This leads to passive and lossless surface designs, which are highly desirable in practice as they do not require active and/or lossy components.

\end{abstract}

\maketitle

\section{Introduction}
\label{sec:intro}
Recent advancements in metasurfaces (MTSs) have unveiled their ability to control various aspects of electromagnetic waves with unprecedented precision and efficiency. They consist of two-dimensional arrangements of custom-designed deeply-subwavelength electromagnetic scatterers called ``meta-atoms"~\cite{Quevedo_Teruel_2019,HMS_review}. By properly tuning the responses of the their constituent meta-atoms, MTSs can be engineered to provide a wide range of functionalities such as wave redirection~\cite{beam_redirection1,Joseph,Paul_AHMS2,DBMG}, frequency filtering~\cite{FSS1,FSS2}, polarization conversion~\cite{Pol_transformer,Min_CPSS}, beam forming~\cite{Michael_Chen_lens,ariel_antenna,Abdo}, and so on.

Due to their low profiles, MTSs are often be modeled as infinitesimally thin sheets of electric and/or magnetic polarization currents which are tailored to enforce certain field discontinuities~\cite{HMS1}, according to appropriate boundary conditions~\cite{GSTCs}. In so-called ``omega-bianisotropic metasurfaces"~\mbox{(O-BMSs)}, which are the main focus of this paper, the induced electric and magnetic currents are cross-coupled. This feature grants them extended functionalities compared to conventional ``Huygens' metasurfaces" (HMSs) in which the coupling is absent~\cite{Suscept_tensor}. The boundary conditions corrseponding to general \mbox{O-BMSs} are the bianisotropic sheet transition conditions (BSTCs)~\cite{local_power_conservation}.

In the past few years, non-planar HMSs as well as \mbox{O-BMSs} have generated a great deal of interest. In particular, cylindrical MTS-based devices with arbitrary cross-sectional shapes have been extensively investigated and employed due to the practical applications made possible by their unique geometries. They are the ideal platform for construction of electromagnetic surface cloaks, since they can be molded at will to offer complete low-profile enclosures for arbitrarily shaped scatterers~\cite{michael_cloak, paris_cloak, Caloz_cloak, lossless_cloak}. Their ability to confine sources of radiation or scattering obstacles has also been leveraged to synthesize electromagnetic illusions~\cite{capolino_illusion,Kwon_Illusion} and high-gain conformal antennas~\cite{cylindrical_antenna}.

Many efficient and accurate numerical analysis techniques for arbitrarily shaped cylindrical MTSs have been developed~\cite{Sandeep_MoM, Caloz_MoM,multifilament}. However, synthesis of such surfaces has been mostly based on the direct solution of the BSTC equations, which generally results in active and/or lossy designs. These features are undesirable because the generation or absorption of power can be hard to realize/control in practice.

Recently, we proposed a versatile modal expansion framework which can be wielded as an analysis as well as a synthesis tool for cylindrical metasurfaces~\cite{xu2020discrete}. Notably, it can guarantee passivity and losslessness of the generated designs, which consequentially are easily  implementable with only reactive components such as etched conductive patterns on printed circuit boards~\cite{angular_momentum}.

In this work, we extend the aforementioned methods to model and design MTSs with irregular cross sections. Due to the geometries of the MTSs which generally are inseparable and thus incompatible with the method of separation of variables, the enforcement of the BSTCs for analytical modal solutions of the Helmholtz wave equation is difficult. One promising approach to address this issue is to use a conformal mapping to transform the physical shape of the MTS into some simple shape, such as a circle, on which the boundary conditions are easily enforced. This technique has been demonstrated to be a viable route for solving scattering~\cite{conformal_acoustic_1,conformal_acoustic_2} and inverse scattering problems~\cite{inverse_scattering} involving homogeneous cylindrical scatterers with irregular cross sections. To the best of our knowledge, spatially varying surface properties such as those possessed by MTSs have not been extensively incorporated into this framework.

By introducing spatially modulated boundary conditions, we unlock previously untapped potentials of the analytical modal expansion technique facilitated by the conformal mapping. Namely, we are able to derive closed-form formulas for the prediction as well as the design of the electromagnetic scattering behaviors of inhomogeneous non-circular cylindrical O-BMSs. More importantly, owing to the well-defined wave impedances associated with the eigenfunctions of the modal expansions, it is easy to identify strictly passive and lossless designs.

To validate the accuracy of the proposed analysis technique, we evaluate the scattered fields from a non-circular cylindrical metasurface and compare the results with those obtained from simulations based on the finite element method (FEM). To demonstrate the effectiveness of the synthesis technique, we design several different metasurface-based devices and confirm their functionalities, also with FEM simulations.

\section{Theory}
\label{sec:theory}
\begin{figure}[b]
\centering
\includegraphics[width=0.9\linewidth]{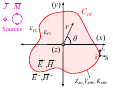}
\caption{Externally excited cylindrical O-BMS with non-circular cross section.}
\label{fig:PhysicalPlane}
\end{figure}

For the problem under consideration, it is natural to describe the physical space occupied by the O-BMS with a two-dimensional polar coordinate system ($r,\theta$), with \mbox{$\partial/\partial z=0$}. This is illustrated in Fig.~\ref{fig:PhysicalPlane}, in which the cross section of the O-BMS is modeled by a simple closed curve $C_{r\theta}$. The metasurface is characterized by its spatially varying scalar electric impedance $Z_{se}$, magnetic admittance $Y_{sm}$ and magnetoelectric coupling $K_{em}$. The materials inside and outside of $C_{r\theta}$ have relative permittivities $\epsilon_{r1}$ and $\epsilon_{r2}$ respectively; they are assumed to be non-magnetic.

Without loss of generality, we assume the electromagnetic field distributions to be transverse magnetic with respect to the $z$-axis ($\mathrm{TM}^z$), with implicit time dependency $e^{j\omega t}$. Thus we can write $\vec{E}=\hat{z}E_z$. Extensions to transverse electric ($\mathrm{TE}^z$) configurations is straightforward and will not be discussed in this work. Furthermore, we assume for now that all (electric and/or magnetic) sources are outside the curve $C_{r\theta}$. After the overall framework is established, it can be trivially extended to accommodate internal sources, as is done in Appendix~\ref{app:Internal_excitation}.

In the source free regions of the physical plane, the electromagnetic fields satisfy the Helmholtz wave equation
\begin{equation}
\label{eqn:Helmholtz}
\nabla^2_{r\theta}\Psi+k_l^2\Psi=0, \quad l\in\{1,2\},
\end{equation}
where $k_l=\sqrt{\epsilon_{rl}}k_o$ is the wavenumber of the host medium. The $\nabla^2_{r\theta}$ operator is the Laplacian in the physical ($r,\theta$) coordinates. The function $\Psi$ can represent any vector component of the electromagnetic fields.

The tangential field discontinuities across the O-BMS boundary is governed by the well-known BSTCs. For the scalar O-BMSs under consideration, this can be stated as:
\begin{equation}
\label{eqn:BSTC}
\begin{split}
\hat{z}\cdot\vec{E}_{av}&=Z_{se}\left(\hat{t}_\parallel\cdot\Delta\vec{H}\right)-K_{em}\left(\hat{z}\cdot\Delta\vec{E}\right),\\
\hat{t}_\parallel\cdot\vec{H}_{av}&=Y_{sm}\left(\hat{z}\cdot\Delta\vec{E}\right)+K_{em}\left(\hat{t}_\parallel\cdot\Delta\vec{H}\right),\\
\end{split}
\end{equation}
where
\begin{equation}
\vec{E}_{av}\triangleq\frac{1}{2}(\vec{E}^++\vec{E}^-),\quad \vec{H}_{av}\triangleq\frac{1}{2}(\vec{H}^++\vec{H}^-)
\end{equation}
are the averaged electromagnetic fields at the boundary, and 
\begin{equation}
\Delta\vec{E}\triangleq \vec{E}^+-\vec{E}^-,\quad \Delta\vec{H}\triangleq \vec{H}^+-\vec{H}^-
\end{equation}
are the field discontinuities. The superscripts  ``$+$" and ``$-$" correspond to the fields immediately outside and inside the metasurface cavity respectively. The vector \mbox{$\hat{t}_\parallel=\hat{r}t_r+\hat{\theta}t_\theta$} is the in-plane tangential unit vector along~$C_{r\theta}$. 

Due to the assumed source location, the tangential fields inside $C_{r\theta}$ consist of the transmitted fields ($E^t_z,H^t_\parallel$), whereas  the external fields consist of the incident ($E^i_z,H^i_\parallel$) plus the reflected fields ($E^r_z,H^r_\parallel$).

To analytically evaluate the scattered fields for any given incident illumination, we must solve (\ref{eqn:Helmholtz}) subject to the boundary \mbox{conditions (\ref{eqn:BSTC})} on $C_{r\theta}$. We attempt to do so with a conformal transformation approach illustrated in Fig.~\ref{fig:PhysicalPlane_and_ComputationPlane}. Under this transformation, the physical plane inhabited by the O-BMS is mapped to a computational plane in which the surface resides on a circular cylinder with radius $\alpha$. This simplifies the enforcement of~(\ref{eqn:BSTC}) on the MTS boundary. In the following subsections, we describe methods to predict (Sec.~\ref{sec:analysis}) and engineer (Sec.~\ref{sec:synthesis}) the electromagnetic behaviors of general cylindrical O-BMSs.

\subsection{Analysis}
\label{sec:analysis}
As a preliminary comment, we acknowledge that it can be difficult to obtain simple analytic mappings which are conformal in the entire complex plane. As a result, the proposed framework generally cannot provide the scattered fields everywhere in space. In this work, we devote our efforts to evaluating the scattered fields only in the exterior region. Although assessment of the interior fields is possible through a separate mapping, it will not be considered in this paper. This is because the effects of most practical cylindrical metasurfaces, including those considered in this paper, are intended to be observed outside of the surface enclosure.

Despite this limitation, it should be noted that the evaluation of the scattered fields immediately interior to the MTS is still possible by virtue of the conformality of the mapping on, and immediately within, $C_{r\theta}$. This has two important implications. First, we can still account for sources located inside the MTS cavity, since we can simply specify the source fields on $C_{r\theta}$. Knowledge of the source fields anywhere else is not required, if we only seek the field distribution outside of $C_{r\theta}$. In Sec.~\ref{sec:results_illusions}, this fact is exploited to design an internally excited electromagnetic illusion MTS. An equally significant corollary is that we are able to engineer the fields inside $C_{r\theta}$ even without access to a proper conformal mapping for that entire bounded region. This is justified by the equivalence principle, because to manipulate the fields in a volume, we simply need to properly engineer the tangential fields on its bounding surface. We use this property to design passive and lossless O-BMS cloaks in Sec.~\ref{sec:results_cloak}.

\begin{figure}[t]
\centering
\includegraphics[width=0.49\textwidth]{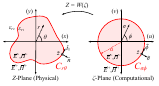}
\caption{Illustration of the conformal mapping between the physical $Z$-plane and the computational $\zeta$-plane.}
\label{fig:PhysicalPlane_and_ComputationPlane}
\end{figure}

To begin our analysis, let us consider a function $W$ which maps a complex variable $\zeta = \sigma e^{j\phi}$ to another variable $Z = re^{j\theta}$, according to
\begin{equation}
\label{eqn:analytic_function}
Z =W(\zeta),
\end{equation}
where ($\sigma,\phi$) are the polar coordinates of the ``computational plane" in which (\ref{eqn:BSTC}) will be enforced. We assume conformality for $W$ in the region $\sigma\geq\alpha$, and that it maps the circle $C_{\sigma\phi}=\{(\sigma,\phi)|\,\sigma=\alpha,\phi\in[0,2\pi]\}$ in the complex $\zeta$-plane to $C_{r\theta}$ in the complex $Z$ plane. Again, this mapping is conceptualized in Fig.~\ref{fig:PhysicalPlane_and_ComputationPlane}.

A well-known result from transformation optics is that under a conformal mapping, the Helmholtz wave equation takes a new form in the $\zeta$-plane given by
\begin{equation}
\label{eqn:conformal_Maxwells_eqn}
\nabla^2_{\sigma\phi}\Psi+\left|\frac{dW}{d\zeta}\right|^2k_l^2\Psi=0,\quad l\in\{1,2\},
\end{equation}
where $\nabla^2_{\sigma\phi}$ is the Laplacian in the ($\sigma,\phi$) coordinates. The term in front of $k_l$ can be interpreted as a spatial modulation acting on the effective refractive index of the host medium~\cite{Favraud2015}. 

General modal solutions to (\ref{eqn:conformal_Maxwells_eqn}) are available and are frequently utilized in acoustic and electromagnetic scattering problems~\cite{conformal_acoustic_1,conformal_acoustic_2}. Following the established results, we can write the incident, reflected and transmitted electric fields as
\begin{equation}
\label{eqn:electric_field_expressions}
\begin{split}
E^i_z &= \sum_{m=-\infty}^{\infty}A^i_mJ_m(k_2\left|W(\zeta)\right|)e^{jm\angle W(\zeta)},\\
E^t_z &= \sum_{m=-\infty}^{\infty}A^t_mJ_m(k_1\left|W(\zeta)\right|)e^{jm\angle W(\zeta)},\\
E^r_z &= \sum_{m=-\infty}^{\infty}A^r_mH_m^{(2)}(k_2\left|W(\zeta)\right|)e^{jm\angle W(\zeta)},
\end{split}
\end{equation}
where $A^{\{i,t,r\}}_m$ are the coefficients corresponding to the $m^{th}$ incident, transmitted and reflect modes. Here, $J_m$ and $H^{(2)}_m$ are the Bessel function and the Hankel function of the second kind respectively.
 
Since the fields are $2\pi$-periodic in $\phi$ for any fixed $\sigma$, we may represent them as $N\times1$ vectors by sampling them at $N$ equally spaced sampling points along the circle $C_{\sigma\phi}$ as follows:

\begin{equation}
\label{eqn:electric_field_vectors}
\begin{split}
\bar{E}^{\{i,r\}}_z[n] &= E^{\{i,r\}}_z(\zeta)\vert_{\sigma\to\alpha^+,\phi = 2(n-1)\pi/N},\\
\bar{E}^t_z[n] &= E^t_z(\zeta)\vert_{\sigma\to\alpha^-,\phi = 2(n-1)\pi/N},\\
n&\in[1,N].
\end{split}
\end{equation}
In general, uniformly distributed sampling points in the $\zeta$-plane will not be mapped to uniformly spaced points in the $Z$-plane. However, this is not necessarily disadvantageous, since we can choose $W$ to yield higher sampling rates at sharp corners in the physical plane, where more rapid variations in surface parameters are expected. 

An equivalent representation of the boundary electric fields is given in terms of their modal expansion coefficients $\hat{A}^{\{i,t,r\}}$ as follows:
\begin{equation}
\label{eqn:modal_expansion_vectors}
\hat{A}^{\{i,t,r\}}[m] = A^{\{i,t,r\}}_{m^\star},\quad m\in [1,M],
\end{equation}
where the shifted index
\begin{equation}
\label{eqn:m_star}
m^\star\triangleq\frac{2m-M-1}{2}
\end{equation}
is introduced so that the center entry of $\hat{A}^{\{i,t,r\}}$ correspond to the fundamental ($0^{th}$) mode. The infinite summations in (\ref{eqn:electric_field_expressions}) have been truncated to $M$ terms to facilitate their computation.

The vectors $\bar{E}^{\{i,t,r\}}_z$ and $\hat{A}^{\{i,t,r\}}$ can be algebraically related to each other by performing a modal decomposition. With the circular cylindrical O-BMSs examined previously, the eigenfunctions of the modal expansions are orthogonal on the physical cross section perimeter. Thus, the modal decomposition of $\bar{E}^{\{i,t,r\}}_z$ amounted to N-point discrete Fourier transforms (N-DFT) which can be implemented using the DFT matrix~\cite{xu2020discrete}. This is not applicable for the present problem, because the modal wavefunctions in (\ref{eqn:electric_field_expressions}) are not orthogonal on $C_{\sigma\phi}$. However, examining the summations, we can still take inspiration from the DFT matrix formalism and write

\begin{equation}
\label{eqn:electric_modal_vectors}
\begin{split}
\bar{E}^{\{i,t,r\}}_z &= \mathbf{P}^{\{i,t,r\}}\hat{A}^{\{i,t,r\}},\\
\mathbf{P}^i[n][m] &= J_{m^\star}(k_2\left|W(\zeta^\star_n)\right|)e^{jm^\star\angle W(\zeta^\star_n)},\\
\mathbf{P}^t[n][m] &= J_{m^\star}(k_1\left|W(\zeta^\star_n)\right|)e^{jm^\star\angle W(\zeta^\star_n)},\\
\mathbf{P}^r[n][m] &= H^{(2)}_{m^\star}(k_2\left|W(\zeta^\star_n)\right|)e^{jm^\star\angle W(\zeta^\star_n)},\\
\end{split}
\end{equation}
where
\begin{equation}
\begin{split}
\label{eqn:zeta_star_n}
\zeta^\star_n&\triangleq\alpha e^{j2(n-1)\pi/N}.\\
\end{split}
\end{equation}

We can derive similar vector representations of the magnetic fields, which can be algebraically related to $\bar{E}^{\{i,t,r\}}$. From (\ref{eqn:electric_field_expressions}), analytical expressions for the magnetic fields may be obtained using time harmonic Maxwell–Faraday equation:
\begin{equation}
\label{eqn:curl_r_theta}
\nabla_{r\theta}\times \vec{E}^{\{i,t,r\}}=-j\omega\mu\vec{H}^{\{i,t,r\}}.
\end{equation}
Under the conformal mapping $W$, equation (\ref{eqn:curl_r_theta}) is transformed into
{\small
\begin{equation}
\renewcommand\arraystretch{1.5}
\label{eqn:curl_sigma_phi}
\frac{1}{h_\sigma h_\phi h_z}\begin{vmatrix}
\hat{\sigma}h_\sigma & \hat{\phi}h_\phi & \hat{z}h_z \\ 
\frac{\partial}{\partial\sigma} & \frac{\partial}{\partial\phi} & \frac{\partial}{\partial z} \\ 
h_\sigma E^{\{i,t,r\}}_\sigma & h_\phi E^{\{i,t,r\}}_\phi & h_zE^{\{i,t,r\}}_z
\end{vmatrix}=-j\omega\mu\vec{H}^{\{i,t,r\}},
\end{equation}}%
where
\begin{equation}
h_\iota = \sqrt{\left|\frac{\partial r}{\partial \iota} \right|^2+\left|\frac{r\partial \theta}{\partial \iota} \right|^2+\left|\frac{\partial z}{\partial \iota} \right|^2}.
\end{equation}
As stated in our initial assumptions, we have $\partial/\partial z = 0$ and $E^{\{i,t,r\}}_\sigma=E^{\{i,t,r\}}_\phi=0$. 

Because $W$ is conformal in our region of interest, the component of $\vec{H}^{\{i,t,r\}}$ tangential to $C_{r\theta}$ corresponds to the $\phi$-component along the circle $C_{\sigma\phi}$. Therefore, to effectively leverage (\ref{eqn:BSTC}), we simply evaluate $H^{\{i,t,r\}}_\phi$, which according to (\ref{eqn:curl_sigma_phi}) are given by:

\begin{figure}[h!]
\begin{equation}
\label{eqn:magnetic_field_expressions}
\begin{split}
H^i_\phi &= \sum_{m=-\infty}^{\infty}\gamma_{2,m}\left(W,\zeta\right)A^i_mJ_m(k_2\left|W\left(\zeta\right)\right|)e^{jm\angle W (\zeta)},\\
H^t_\phi &= \sum_{m=-\infty}^{\infty}\gamma_{1,m}\left(W,\zeta\right)A^t_mJ_m(k_1\left|W\left(\zeta\right)\right|)e^{jm\angle W (\zeta)},\\
H^r_\phi &= \sum_{m=-\infty}^{\infty}\tau_{2,m}\left(W,\zeta\right)A^r_mH^{(2)}_m(k_2\left|W\left(\zeta\right)\right|)e^{jm\angle W (\zeta)},\\
\end{split}
\end{equation}
\end{figure}
where the function compositions
\begin{equation}
\label{eqn:modal_admittance_internal}
\begin{split}
\gamma_{l,m}\left(W,\zeta\right)=&\frac{k_l}{j\omega\mu h_\sigma}\frac{J'_m(k_l\left|W\left(\zeta\right)\right|)}{J_m(k_l\left|W\left(\zeta\right)\right|)}\cdot\frac{\partial}{\partial\sigma}\left|W\left(\zeta\right)\right|\\
&+\frac{m}{\omega\mu h_\sigma}\cdot\frac{\partial}{\partial\sigma}\angle W (\zeta),\\
\tau_{l,m}\left(W,\zeta\right)=&\frac{k_l}{j\omega\mu h_\sigma}\frac{H^{(2)'}_m(k_l\left|W\left(\zeta\right)\right|)}{H^{(2)}_m(k_l\left|W\left(\zeta\right)\right|)}\cdot\frac{\partial}{\partial\sigma}\left|W\left(\zeta\right)\right|\\
&+\frac{m}{\omega\mu h_\sigma}\cdot\frac{\partial}{\partial\sigma}\angle W (\zeta)\\
\end{split}
\end{equation}
can be interpreted as the spatially varying modal wave admittances for various components of the $m^{th}$ mode. The implicit dependency of $h_\sigma$ on $W$ and $\zeta$ is suppressed.

Sampling the magnetic fields above and below the O-BMS at $N$ equally spaced sampling points along the circle $C_{\sigma\phi}$ yields the boundary field vectors

\begin{equation}
\label{eqn:magnetic_field_vectors}
\begin{split}
\bar{H}^{\{i,r\}}_\phi[n] &= H^{\{i,r\}}_z(\zeta)\vert_{\sigma\to\alpha^+,\phi = 2(n-1)\pi/N},\\
\bar{H}^t_\phi[n] &= H^t_z(\zeta)\vert_{\sigma\to\alpha^-,\phi = 2(n-1)\pi/N}.\\
\end{split}
\end{equation}
Alternatively, based on (\ref{eqn:magnetic_field_expressions}), these vectors can be rewritten as
\begin{equation}
\label{eqn:magnetic_modal_vectors}
\begin{split}
\bar{H}^{\{i,t,r\}}_\phi =& \mathbf{Q}^{\{i,t,r\}}\hat{A}^{\{i,t,r\}},\\
\mathbf{Q}^i[n][m] =&\gamma_{2,m^\star}\left(W,\zeta^\star_n\right)\cdot J_{m^\star}\left(k_2\left|W(\zeta^\star_n)\right|\right)e^{jm^\star\angle W(\zeta^\star_n)},\\
\mathbf{Q}^t[n][m] =&\gamma_{1,m^\star}\left(W,\zeta^\star_n\right)\cdot J_{m^\star}\left(k_1\left|W(\zeta^\star_n)\right|\right)e^{jm^\star\angle W(\zeta^\star_n)},\\
\mathbf{Q}^r[n][m] =&\tau_{2,m^\star}\left(W,\zeta^\star_n\right)\cdot H^{(2)}_{m^\star}\left(k_2\left|W(\zeta^\star_n)\right|\right)e^{jm^\star\angle W(\zeta^\star_n)},
\end{split}
\end{equation}
where $m^\star$ and $\zeta^\star_n$ are as defined in (\ref{eqn:m_star}) and (\ref{eqn:zeta_star_n}) respectively.

Comparing (\ref{eqn:electric_modal_vectors}) with (\ref{eqn:magnetic_modal_vectors}) leads to the algebraic relationship between the boundary electric and the magnetic field vectors as follows:

\begin{equation}
\label{eqn:electric_magnetic_field_relations}
\begin{split}
\bar{H}^{\{i,t,r\}}_\phi &= \mathbf{Y}^{\{i,t,r\}}\bar{E}^{\{i,t,r\}}_z,\\
\mathbf{Y}^{\{i,t,r\}} &\triangleq \mathbf{Q}^{\{i,t,r\}}\left(\mathbf{P}^{\{i,t,r\}}\right)^{-1}.\\
\end{split}
\end{equation}

The matrices $\mathbf{Y}^{\{i,t,r\}}$ can be interpreted as admittance matrices which relate the incident, transmitted and reflected electric and magnetic fields for externally excited metasurfaces. 

Having obtained vector representations of the boundary electromagnetic fields, we can also sample the O-BMS surface parameters at the same sampling points, obtaining $N\times1$ surface property vectors $\bar{Z}_{se}$, $\bar{Y}_{sm}$ and $\bar{K}_{em}$. This allows us to write the discretized form of the BSTC equations as
\begin{equation}
\label{eqn:BSTC_discrete_analysis}
\begin{split}
&\frac{1}{2}\left(\bar{E}^t_z+\bar{E}^i_z+\bar{E}^r_z\right)\\
=&\mathbf{Z}\left(\mathbf{Y}^t\bar{E}^t_z-\mathbf{Y}^i\bar{E}^i_z-\mathbf{Y}^r\bar{E}^r_z\right)-\mathbf{K}\left(\bar{E}^t_z-\bar{E}^i_z-\bar{E}^r_z\right),\\
&\frac{1}{2}\left(\mathbf{Y}^t\bar{E}^t_z+\mathbf{Y}^i\bar{E}^i_z+\mathbf{Y}^r\bar{E}^r_z\right)\\
=&\mathbf{Y}\left(\bar{E}^t_z-\bar{E}^i_z-\bar{E}^r_z\right)+\mathbf{K}\left(\mathbf{Y}^t\bar{E}^t_z-\mathbf{Y}^i\bar{E}^i_z-\mathbf{Y}^r\bar{E}^r_z\right),\\
\end{split}
\end{equation}
where
\begin{equation}
\mathbf{Z}=\mathrm{diag}(\bar{Z}_{se}),\,\, \mathbf{Y}=\mathrm{diag}(\bar{Y}_{sm}),\,\,\mathbf{K}=\mathrm{diag}(\bar{K}_{em}).
\end{equation}
Note that the boundary magnetic field vectors are absent from (\ref{eqn:BSTC_discrete_analysis}), owing to the application of (\ref{eqn:electric_magnetic_field_relations}), which eliminates half of the unknowns in the analysis problem. Following the same idea presented previously for circular cylindrical O-BMSs, we can then rearrange (\ref{eqn:BSTC_discrete_analysis}) to derive closed-form expressions for the transmitted and reflected boundary electric fields. They can be written in terms of electric field transmission ($\mathbf{T}_e$) and reflection ($\mathbf{R}_e$) matrices, according to
\begin{equation}
\label{eqn:TR_definition}
\bar{E}^t_z = \mathbf{T}_e\bar{E}^i_z,\,\,\bar{E}^r_z = \mathbf{R}_e\bar{E}^i_z.
\end{equation}
The explicit expressions for these matrices are given by
\begin{equation}
\label{eqn:T_R_ext}
\begin{split}
\mathbf{T}_e = &\mathbf{t}_a^{-1}\mathbf{t}_b,\quad\mathbf{R}_e = \mathbf{r}_a^{-1}\mathbf{r}_b,\\
\mathbf{t}_a = & \left(\frac{1}{2}\mathbf{Y}^r-\mathbf{Y}-\mathbf{K}\mathbf{Y}^r\right)^{-1}\left(\frac{1}{2}\mathbf{Y}^t+\mathbf{Y}+\mathbf{K}\mathbf{Y}^t\right),\\
     &-\left(\frac{1}{2}\mathbf{I}-\mathbf{Z}\mathbf{Y}^r+\mathbf{K}\right)^{-1}\left(\frac{1}{2}\mathbf{I}+\mathbf{Z}\mathbf{Y}^t-\mathbf{K}\right)\\
\mathbf{t}_b = & \left(\frac{1}{2}\mathbf{I}-\mathbf{Z}\mathbf{Y}^r+\mathbf{K}\right)^{-1}\left(\frac{1}{2}\mathbf{I}-\mathbf{Z}\mathbf{Y}^i+\mathbf{K}\right)\\
     &-\left(\frac{1}{2}\mathbf{Y}^r-\mathbf{Y}-\mathbf{K}\mathbf{Y}^r\right)^{-1}\left(\frac{1}{2}\mathbf{Y}^i-\mathbf{Y}-\mathbf{K}\mathbf{Y}^i\right),\\
\mathbf{r}_a = &\left(\frac{1}{2}\mathbf{I}+\mathbf{Z}\mathbf{Y}^t-\mathbf{K}\right)^{-1}\left(\frac{1}{2}\mathbf{I}-\mathbf{Z}\mathbf{Y}^r+\mathbf{K}\right)\\
								&-\left(\frac{1}{2}\mathbf{Y}^t+\mathbf{Y}+\mathbf{K}\mathbf{Y}^t\right)^{-1}\left(\frac{1}{2}\mathbf{Y}^r-\mathbf{Y}-\mathbf{K}\mathbf{Y}^r\right),\\ 			
\mathbf{r}_b = & \left(\frac{1}{2}\mathbf{Y}^t+\mathbf{Y}+\mathbf{K}\mathbf{Y}^t\right)^{-1}\left(\frac{1}{2}\mathbf{Y}^i-\mathbf{Y}-\mathbf{K}\mathbf{Y}^i\right)\\
     &-\left(\frac{1}{2}\mathbf{I}+\mathbf{Z}\mathbf{Y}^t-\mathbf{K}\right)^{-1}\left(\frac{1}{2}\mathbf{I}-\mathbf{Z}\mathbf{Y}^i+\mathbf{K}\right).\\
\end{split}
\end{equation}
Equivalently, we can write
\begin{equation}
\label{eqn:TR_definition_alt}
\begin{split}
\hat{A}^t &= \left(\mathbf{P}^t\right)^{-1}\mathbf{T}_e\mathbf{P}^i\hat{A}^i\triangleq\mathbf{T}\hat{A}^i,\\
\hat{A}^r &= \left(\mathbf{P}^r\right)^{-1}\mathbf{R}_e\mathbf{P}^i\hat{A}^i\triangleq\mathbf{R}\hat{A}^i,
\end{split}
\end{equation}
where $\mathbf{T}$ and $\mathbf{R}$ are the modal transmission and reflection matrices. Using (\ref{eqn:T_R_ext}) and (\ref{eqn:TR_definition_alt}), we can evaluate the $\hat{A}^{\{t,r\}}$ corresponding to any set of given $\{\bar{Z}_{se},\bar{Y}_{sm},\bar{K}_{em},\hat{A}^i\}$. Then, the electromagnetic fields everywhere for which $W$ is conformal can be computed using truncated versions of (\ref{eqn:electric_field_expressions}) and (\ref{eqn:magnetic_field_expressions}).

\subsection{Synthesis}
\label{sec:synthesis}
As discussed in Sec.~\ref{sec:intro}, for practical reasons, we are often interested in O-BMSs which can be implemented using only reactive components. Thus, although the preceding analysis is valid for general surfaces which may generate and/or absorb power, we will restrict further investigations to those that are passive and lossless.

A straightforward method of guaranteeing passivity and losslessness is to enforce the condition of local power conservation (LPC), which is essentially a point-wise continuity condition for the normal flux of the real power across the metasurface. In the computational domain, it can be stated as
\begin{equation}
\label{eqn:LPC}
\mathrm{Re}\{(\bar{E}_z^i+\bar{E}_z^r)\odot(\bar{H}_\phi^i+\bar{H}_\phi^r)^*-\bar{E}_z^t\odot(\bar{H}^t_\phi)^*\} = 0,
\end{equation}
where $\odot$ denotes element-wise product and $(\cdot)^*$ denotes complex conjugation.

Previously, it was demonstrated that transmissive scalar metasurfaces can be made passive and lossless if it is made to support some judiciously designed auxiliary reflection, while reflective surfaces can achieve this with some auxiliary transmission~\cite{xu2020discrete}. However, identification of the unknown auxiliary fields is generally non-trivial, since the required auxiliary electric fields and auxiliary magnetic fields must satisfy Maxwell's equations in addition to (\ref{eqn:LPC}). In the present scenario, the problem is compounded by the non-separable geometry of the metasurface, which renders the approach of brute force optimization on the electromagnetic field distribution~\cite{lossless_cloak,Kwon_Illusion} exceedingly difficult.

With the modal solutions obtained in Sec.~\ref{sec:analysis}, the total number of unknowns in the LPC equation is halved. More specifically, using (\ref{eqn:electric_modal_vectors}) and (\ref{eqn:magnetic_modal_vectors}), we can transform (\ref{eqn:LPC}) into

\begin{equation}
\label{eqn:LPC_T}
\mathrm{Re}\{(\bar{E}^i+\mathbf{P}^r\hat{A}^r)\odot(\bar{H}^i+\mathbf{Q}^r\hat{A}^r)^*-\bar{E}^t\odot(\bar{H}^{t})^*\} = 0
\end{equation}
for transmissive O-BMSs, and
\begin{equation}
\label{eqn:LPC_R}
\mathrm{Re}\{(\bar{E}^i+\bar{E}^r)\odot(\bar{H}^i+\bar{H}^r)^*-\mathbf{P}^t\hat{A}^t\odot(\mathbf{Q}^t\hat{A}^t)^*\} = 0
\end{equation}
for reflective O-BMSs. Evaluation of the required auxiliary fields amounts to solving (\ref{eqn:LPC_T}) for the unknown $\hat{A}^r$, or solving (\ref{eqn:LPC_R}) for the unknown $\hat{A}^t$. Either case will not be challenging since these are straightforward non-linear algebraic equations.

If the field transformations of an O-BMS satisfies LPC, then its constituent surface parameters have closed-form expressions given by~\cite{local_power_conservation}:
\begin{equation}
\label{eqn:synthesis}
\begin{split}
\bar{K}_{em} = &\frac{\mathrm{Re}\left\{\bar{E}_z^+\odot\bar{H}_\phi^{-*}-\bar{E}_z^-\odot\bar{H}_\phi^{+*}\right\}}{2\mathrm{Re}\left\{\left(\bar{E}_z^+-\bar{E}_z^-\right)\odot\left(\bar{H}_\phi^+-\bar{H}_\phi^-\right)^*\right\}},\\
\bar{Y}_{sm} = &\frac{j}{2}\mathrm{Im}\left\{\frac{\bar{H}_\phi^++\bar{H}_\phi^-}{\bar{E}_z^+-\bar{E}_z^-}\right\}-j\bar{K}_{em}\odot\mathrm{Im}\left\{\frac{\bar{H}_\phi^+-\bar{H}_\phi^-}{\bar{E}_z^+-\bar{E}_z^-}\right\},\\
\bar{Z}_{se} = &\frac{j}{2}\mathrm{Im}\left\{\frac{\bar{E}_z^++\bar{E}_z^-}{\bar{H}_\phi^+-\bar{H}_\phi^-}\right\}+j\bar{K}_{em}\odot\mathrm{Im}\left\{\frac{\bar{E}_z^+-\bar{E}_z^-}{\bar{H}_\phi^+-\bar{H}_\phi^-}\right\},\\
\end{split}
\end{equation}
where
\begin{equation}
\label{eqn:fields_ext_excited}
\begin{split}
\bar{E}^+_z = \bar{E}^i_z + \bar{E}^r_z,\quad \bar{E}^-_z = \bar{E}^t_z,\\
\bar{H}^+_\phi = \bar{H}^i_\phi + \bar{H}^r_\phi,\quad \bar{H}^-_\phi = \bar{H}^t_\phi,\\
\end{split}
\end{equation}
since we have assumed external sources.

To summarize, the proposed design method for passive and lossless conformal cylindrical O-BMSs has three main steps. First, $\{\mathbf{P}^r,\mathbf{Q}^r\}$ or $\{\mathbf{P}^t,\mathbf{Q}^t\}$ are populated using (\ref{eqn:electric_modal_vectors}) and (\ref{eqn:magnetic_modal_vectors}). Then, (\ref{eqn:LPC_T}) or (\ref{eqn:LPC_R}) is solved to obtain the required auxiliary $\hat{A}^r$ or $\hat{A}^t$. Lastly, with knowledge of the complete field distributions at the MTS boundary, (\ref{eqn:synthesis}) is used to derive the required O-BMS surface parameters.

\subsection{Conformal Transformations}
\label{sec:conformal_map}
For a given cross-sectional shape, we must identify the appropriate function $W$ which maps it to a circle. Although general transformations based on series expansions are available~\cite{conformal_acoustic_1}, many practical geometries do not require such a comprehensive description. In this work, as a proof of concept, we consider a simple function

\begin{equation}
\label{eqn:conformal_map}
\begin{split}
Z = W(\zeta)= R\left(\zeta+\frac{1}{q\zeta^q}\right)e^{jt},\quad q\in\mathbb{Z}\\
\end{split}
\end{equation}
which maps the exterior of a circle centered at the origin of the $\zeta$-plane to the exteriors of various hypotrochoids in the $Z$-plane~\cite{conformal_map_book}. For $q=1$, it can be used to model ellipses. For $q>1$, the resulting hypotrochoids resemble equilateral polygons with $(q+1)$ edges joined by rounded corners. The parameters $R$ and $t$ stretch and rotate the mapped shape in the $Z$-plane respectively. Other metasurface configurations that require more sophisticated mappings are reserved for future investigations.

\subsection{Numerical Considerations}
\label{sec:numerical_considerations}
Modal expansion methods involving Hankel functions are known to suffer from numerical issues when they are used to compute the scattered fields from obstacles with large aspect ratios, such as highly elongated ellipses. This is due to the high amplitude and the highly oscillatory nature of their imaginary components for small radial arguments, which demands high numerical precision to be properly evaluated. 

This problem has been extensively investigated in the context of the transition matrix (T-matrix) method for electromagnetic and acoustic scattering calculations~\cite{T_matrix_problem1,T_matrix_problem2}. Many simple solutions are available. For instance, one can expand the fields with alternative eigenfunctions based on the problem. In this work, we take a more general approach by increasing the numerical precision of our calculations~\cite{multiprecision_toolbox}.

\section{Results and Discussions}
\label{sec:results}

In the following subsections, two-dimensional finite element simulations are conducted with COMSOL Multiphysics to provide validations for the proposed analysis and synthesis techniques. In COMSOL, the O-BMSs are modeled as electric and magnetic polarization current sheets whose complex amplitudes depend on the local averaged fields $\vec{E}_{av}$ and $\vec{H}_{av}$, as described by:

\begin{equation}
\label{eqn:COMSOL_currents}
\begin{split}
\vec{J}_s &= \hat{z}\left[\frac{j\mathrm{Im}\{Y_{sm}\}\left(\hat{z}\cdot\vec{E}_{av}\right)+\mathrm{Re}\{K_{em}\}\left(\hat{t}_\parallel\cdot\vec{H}_{av}\right)}{\mathrm{Re}\{K_{em}\}^2-\mathrm{Im}\{Y_{sm}\}\mathrm{Im}\{Z_{se}\}}\right],\\ 
\vec{M}_s &= \hat{t}_\parallel\left[\frac{j\mathrm{Im}\{Z_{se}\}\left(\hat{t}_\parallel\cdot\vec{H}_{av}\right)-\mathrm{Re}\{K_{em}\}\left(\hat{z}\cdot\vec{E}_{av}\right)}{\mathrm{Re}\{K_{em}\}^2-\mathrm{Im}\{Y_{sm}\}\mathrm{Im}\{Z_{se}\}}\right],\\
\end{split}
\end{equation}
where $\vec{J}_s$ and $\vec{M}_s$ are the surface electric and magnetic current densities respectively.

It is easy to see that the currents specified by (\ref{eqn:COMSOL_currents}) correspond to those induced in a passive and lossless O-BMS, since the real parts of $\bar{Z}_{se}$ and $\bar{Y}_{sm}$ and the imaginary parts of $\bar{K}_{em}$, which are responsible for power gain and/or dissipation~\cite{passive_lossless_parameter}, are identically zero.

In COMSOL, the discrete vectors $\bar{Z}_{se}$, $\bar{Y}_{sm}$ and $\bar{K}_{em}$ are interpolated using rectangular pulses to form continuous surfaces. Conveniently, the local tangential unit vectors $\hat{t}_\parallel$ at any point along a parametric curve is readily available as a built-in variable in COMSOL. 

Lastly, we note that in COMSOL, cross-sectional shapes with high aspect ratios may require dense FEM meshes in order to properly resolve their fine features. Designs with rapidly varying surface properties also demand careful meshing. Nevertheless, all the designs presented in this paper have short simulation times on the order of minutes.

\subsection{Evaluation of Scattered Fields}
\label{sec:results_scattered_fields}
As a validation of the proposed framework, we first calculate the external scattered fields of a quasi-square dielectric cylinder coated with an impedance surface characterized by its homogeneous properties 
\begin{equation}
	Z_{se}=j100,\quad Y_{sm}\to\infty, \quad K_{em}=0.
\end{equation}
The incident field is a $z$-polarized cylindrical wave radiated by an external electric line source located at $(r,\theta)=(r_s,\theta_s)$; its modal expansion can be inferred from the addition theorem for Hankel functions to be~\cite{balanis}:
\begin{equation}
\label{eqn:external_linesource_Ei}
\hat{A}^i[m] = H^{(2)}_{m^\star}(k_2r_s)e^{-jm^\star\theta_s}. 
\end{equation}

\begin{figure}[b]
\centering  
\includegraphics[width=0.98\linewidth]{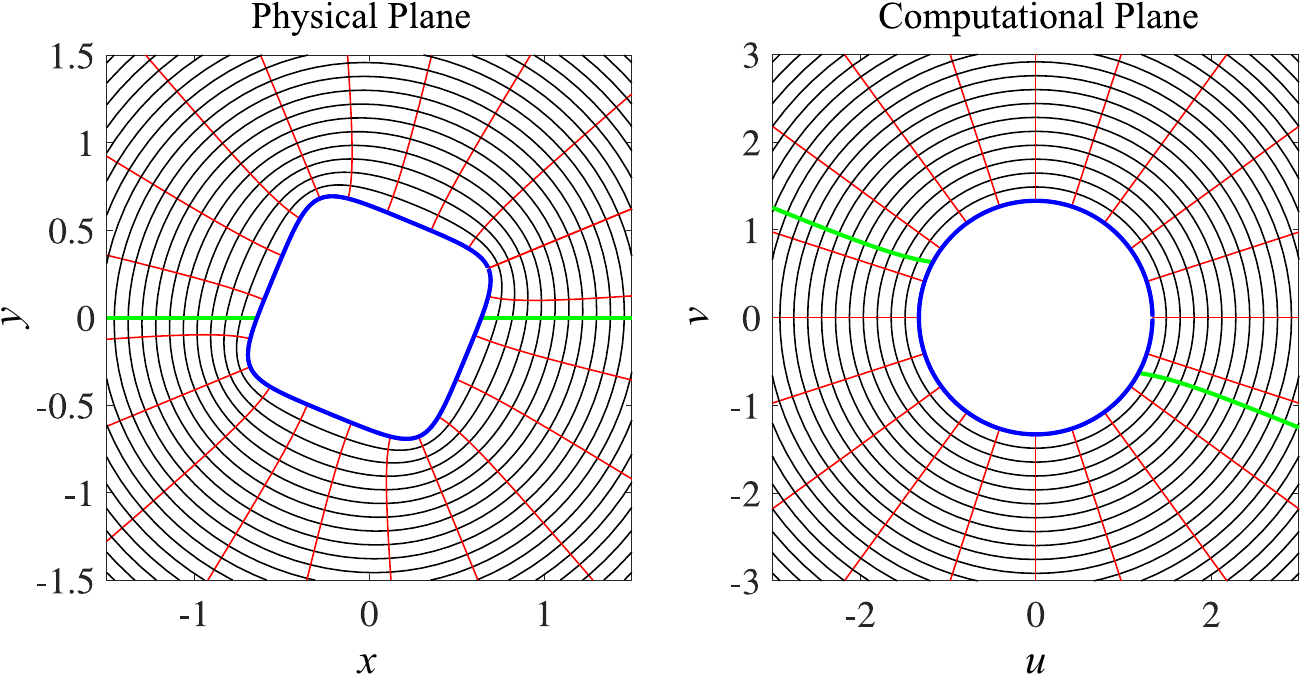}
\caption{Visualization of the conformal transformation specified in Tab.~\ref{tab:ex_A_spec}.}
\label{fig:ex_A_conformal_visualization}
\end{figure}

The relevant geometric and material parameters for this problem are summarized in Tab.~\ref{tab:ex_A_spec}. The corresponding conformal transformation for the external region is visualized in Fig.~\ref{fig:ex_A_conformal_visualization}. The red and black contours correspond to loci of constant $\phi$ and constant $\sigma$ respectively. The dielectric interface and the cross section of the impedance surface in both planes are highlighted with the blue curves.

\begin{table}[h]
\begin{ruledtabular}
\caption{Parameters for the quasi-square cylinder studied in Example A.}
\label{tab:ex_A_spec}
\begin{tabular}{ccccccccccc}
  $f$ (GHz) & $N$ & M &R&q&t& $\alpha$&$\epsilon_{r1}$&$\epsilon_{r2}$ &$r_s$(m)&$\theta_s$\\ 
\hline
 1 & 351 & 251 & 0.5&3&$\pi/8$&1.33 & 3 &1 &2.8&0\\ 
\end{tabular}
\end{ruledtabular}
\end{table}

\begin{figure}[ht!]
\centering  
\includegraphics[width=0.98\linewidth]{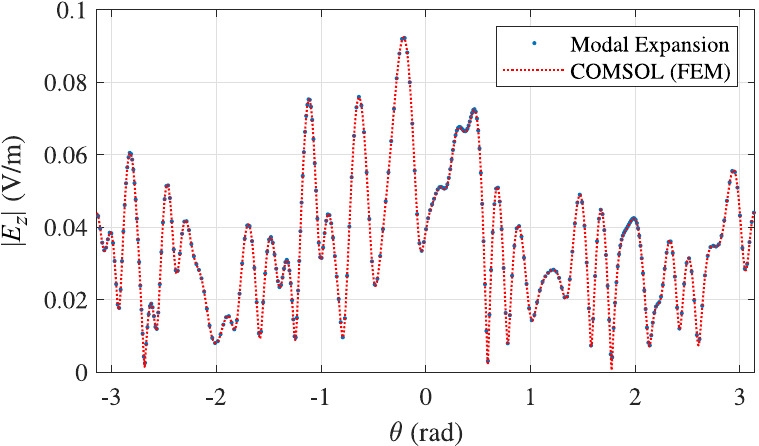}
\caption{Magnitude of the computed total electric fields along the impedance surface (blue curves in Fig.~\ref{fig:ex_A_conformal_visualization}).}
\label{fig:ex_A_MTS_fields}
\end{figure}

In the following plots, we show comparison between the fields evaluated using (\ref{eqn:TR_definition_alt}) and those obtained from COMSOL. First, we calculate the average electric fields on the surface (along the blue curves); the results are plotted in Fig.~\ref{fig:ex_A_MTS_fields}. The horizontal axis corresponds to the physical angular coordinate $\theta$, which has a one-to-one correspondence with the computational angular coordinate $\phi$ for the present geometry.

\begin{figure}[ht!]
\centering  
\includegraphics[width=0.98\linewidth]{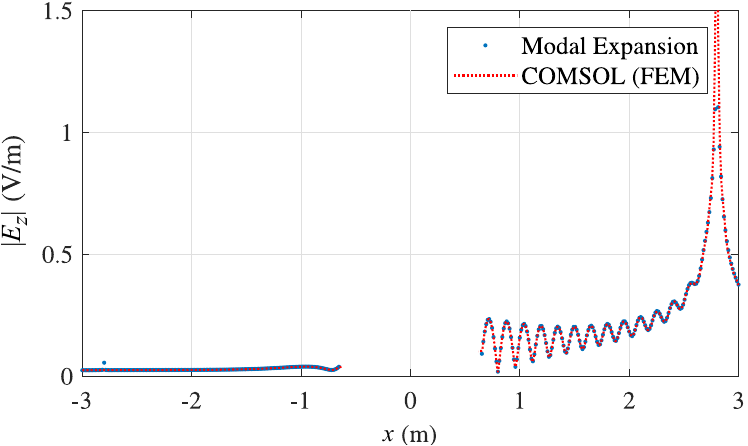}
\caption{Magnitude of the computed total electric fields along the line $y=0$ (green curves in Fig.~\ref{fig:ex_A_conformal_visualization}).}
\label{fig:ex_A_line_fields}
\end{figure}

In Fig.~\ref{fig:ex_A_line_fields}, we plot the total external electric field distribution along the line $y=0$, which is marked as the green curves in Fig.~\ref{fig:ex_A_conformal_visualization}.

In both of the cases examined above, we observe an almost perfect match between the modal expansion prediction and FEM results. The agreement in Fig.~\ref{fig:ex_A_MTS_fields} is particularly encouraging because it lends credence to the proposed framework as an analysis tool for the boundary fields. As we know, the tangential field profile at a metasurface boundary defines its very functionality.

\subsection{Conformal O-BMS Cloaks}
\label{sec:results_cloak}
In this section, we use the proposed synthesis technique to design passive and lossless electromagnetic cloaks for penetrable (dielectric) and impenetrable (perfect conductor) cylindrical objects with non-circular cross sections and arbitrary electrical sizes.

\subsubsection{Penetrable Objects}
\label{sec:penetrable_targets}
First, we design a penetrable O-BMS cloak for dielectric cylinders. Recently, penetrable cloaks have been proposed as a passive and lossless low-profile solution for concealment of certain objects from known external illuminations~\cite{Caloz_cloak,xu2020discrete}. It was shown that if the object permits internal fields, then an O-BMS enclosure can be designed to satisfy LPC by matching the power flow profiles of the incident and transmitted fields. At the same time, the surface can be tailored to incur zero reflection, rendering the target invisible to all external observers.

Penetrable cloaks possess several advantages and disadvantages when compared to other cloaking solutions. For instance, volumetric metamaterial cloaks based on transformation optics can hide objects of any electrical size without \emph{a priori} knowledge of the incident field~\cite{MTM_cloak}, but suffer from complexity, bulkiness and increased loss. Thin plasmonic mantle cloaks can be used to dramatically suppress the dominant component of the scattered fields from electrically small objects~\cite{mantle_cloak_1,mantle_cloak_2}; they perform ideally within the quasistatic regime. Active Huygens' metasurface cloaks can conceal any object from arbitrary external illuminations by virtue of their reconfigurability~\cite{michael_cloak,paris_cloak}, but demand constant power supply and may suffer from stability issues caused by their complex control circuitries. Penetrable cloaks forfeit the robustness afforded by prior designs, since they require knowledge of the external fields. By doing so, they gain substantial improvements in terms of the efficiency, manufacturability and integrability. They are ideal for applications such as electromagnetic interference reduction inside wireless communication systems~\cite{EMC_cloak, Decoupling_Cloak}.

\begin{figure}[t]
\centering  
\includegraphics[width=0.98\linewidth]{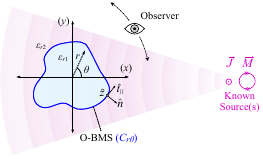}
\caption{Conformal scalar O-BMS cloak for penetrable dielectric objects.}
\label{fig:penetrable_conformal_cloak_schematic}
\end{figure}

The general schematic for a penetrable O-BMS cloak for a non-circular dielectric cylinder is shown in Fig.~\ref{fig:penetrable_conformal_cloak_schematic}. For illustrative purposes, we consider a cylinder with an elliptic cross section and assume the external incident fields to be those radiated by a $\hat{z}$-directed line source located at ($r,\theta$)=($r_s,\theta_s$). The detailed geometric and material parameters are outlined in Tab.~\ref{tab:ex_B1_spec}. With it, we populate all relevant matrices described in Sec.~\ref{sec:theory}.

\begin{table}[b]
\begin{ruledtabular}
\caption{Parameters for the elliptic dielectric cylinder to be concealed.}
\label{tab:ex_B1_spec}
\begin{tabular}{ccccccccccc}
  $f$ (GHz) & $N$ & M &R&q&t&$\alpha$& $\epsilon_{r1}$& $\epsilon_{r2}$ &$r_s$(m)&$\theta_s$\\ 
\hline
 1 & 351& 101& 1/3& 1& $\pi$/4 &1.3& 3& 1&1.33&0\\ 
\end{tabular}
\end{ruledtabular}
\end{table}

\begin{figure}[t]
\centering  
\includegraphics[width=0.98\linewidth]{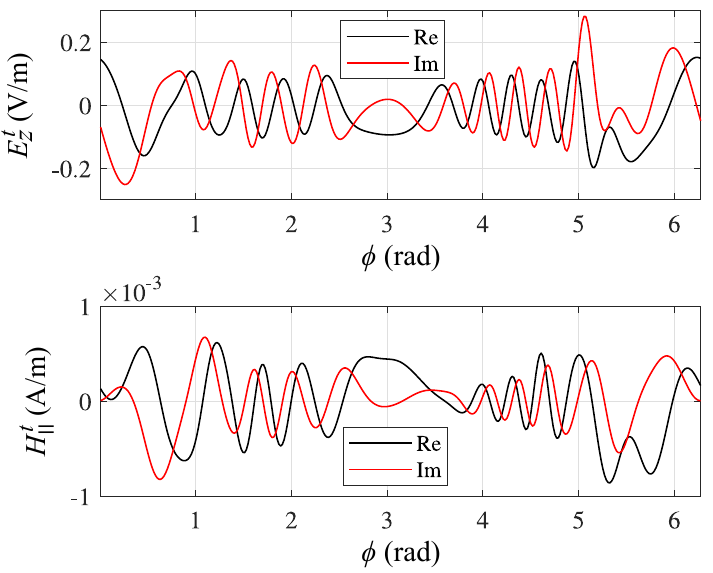}
\caption{The required transmitted auxiliary electric field $E^t_z$ and magnetic field $H^t_\parallel$($\equiv H^t_\phi$) for the passive and lossless cloaking of the elliptic dielectric cylinder.}
\label{fig:elliptic_dielectric_aux_field}
\end{figure}

Next, we note that the penetrable cloak is a reflective MTS, since we are specifying the desired reflection to be zero. Thus, to ensure its passivity and losslessness, we solve the reflective LPC equation (\ref{eqn:LPC_R}) for the unknown auxiliary $\hat{A}^t$, with $\hat{A}^i$ given by (\ref{eqn:external_linesource_Ei}) and $\hat{A}^r=0$. Using the $fsolve$ function in MATLAB, with the Levenberg-Marquardt algorithm, we obtain the required auxiliary $\hat{A}^t$, whose associated boundary electric and magnetic fields are plotted in Fig.~\ref{fig:elliptic_dielectric_aux_field}. Here, the solved vectors $\bar{E}^t_z$ and $\bar{H}^t_\parallel$ are interpolated to form continuous functions of the computational angular coordinate $\phi$.

\begin{figure}[b]
\centering  
\includegraphics[width=0.98\linewidth]{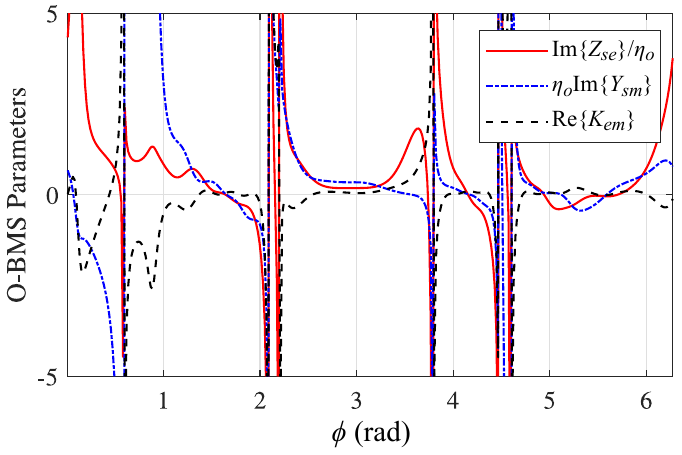}
\caption{Synthesized passive and lossless O-BMS parameters for cloaking the elliptic dielectric cylinder from an external line source.}
\label{fig:elliptic_dielectric_OBMS}
\end{figure}

As the final step, to synthesize the required O-BMS parameters, we insert the complete electromagnetic field distribution into (\ref{eqn:synthesis}). The resultant $\{\bar{Z}_{se}, \bar{Y}_{sm}, \bar{K}_{em}\}$ are interpolated and plotted in Fig.~\ref{fig:elliptic_dielectric_OBMS}. For legibility, $Z_{se}$ and $Y_{sm}$ are normalized against the free space wave impedance $\eta_o$. We note that the real parts of $Z_{se}$ and $Y_{sm}$ as well as the imaginary part of $K_{em}$ are excluded from the plot, since they are identically zero. 

\begin{figure}[ht!]
\centering  
\subfigure[]{\includegraphics[width=0.98\linewidth]{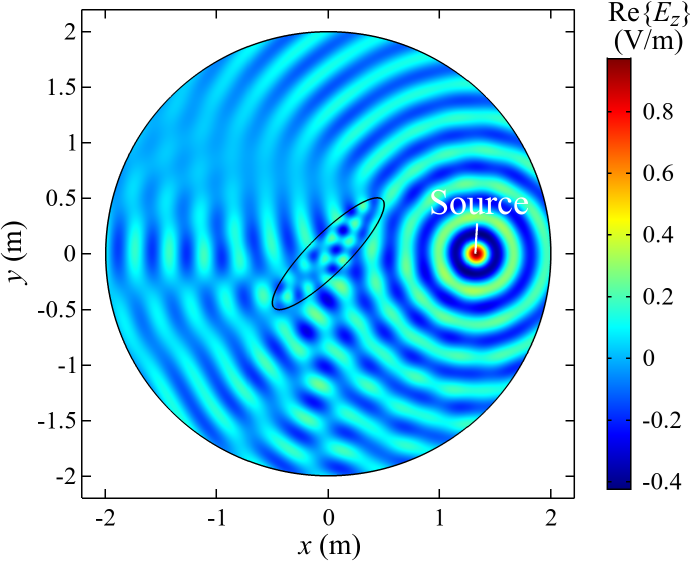}}
\\~\\
\subfigure[]{\includegraphics[width=0.98\linewidth]{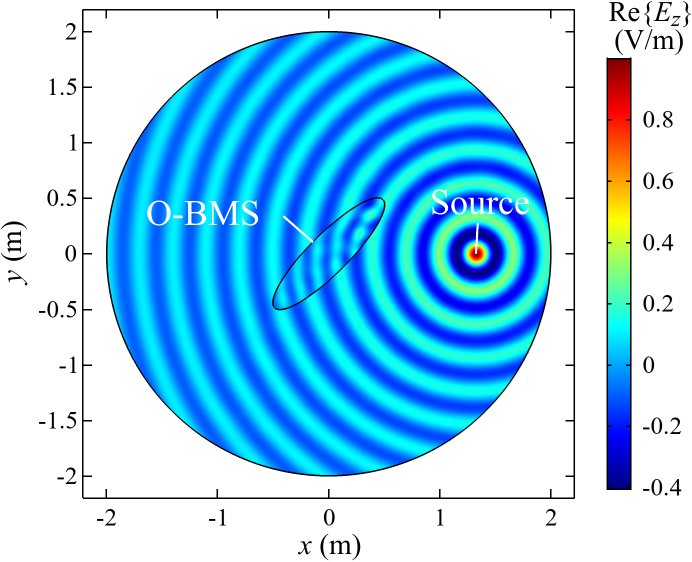}}
\caption{Snapshot of simulated $\mathrm{Re}\{E_z\}$ for an elliptic dielectric cylinder illuminated by an electric line source (a) without  cloaking; and (b) with the O-BMS cloak.}
\label{fig:penetrable_cloak_COMSOL}
\end{figure}

We implement the surface parameters of Fig.~\ref{fig:elliptic_dielectric_OBMS} in COMSOL Multiphysics following (\ref{eqn:COMSOL_currents}) and illuminate the cylinder with the cylindrical source. A snapshot of the resulting electric field distribution $\mathrm{Re}\{E_z\}$ is plotted in Fig.~\ref{fig:penetrable_cloak_COMSOL}(b). The wavefronts emanating from the source location are unperturbed by the cloaked target, signifying no observable reflections. For reference, the field distribution without the O-BMS cloak is shown in  Fig.~\ref{fig:penetrable_cloak_COMSOL}(a). Here, the external reflections caused by the dielectric cylinder create a complex interference pattern in conjunction with the incident cylindrical wave.

\subsubsection{Impenetrable Objects}
\label{sec:impenetrable_targets}
The cloaking configuration shown in Fig.~\ref{fig:penetrable_conformal_cloak_schematic} is unsuitable for impenetrable targets such as perfect electric conductors (PECs), for which $\hat{A}^t=0$. This is because it would be generally impossible to satisfy (\ref{eqn:LPC_R}) while also demanding $\hat{A}^r=0$. Tensorial surface cloaks which leverage orthogonally polarized surface waves to achieve local power balance have been proposed as a potential solution~\cite{lossless_cloak}. However, their constituent meta-atoms may be difficult to realize in practice, since they require intricate anisotropic geometries to perform accurate polarization manipulation. 

\begin{figure}[t]
\centering  
\includegraphics[width=0.98\linewidth]{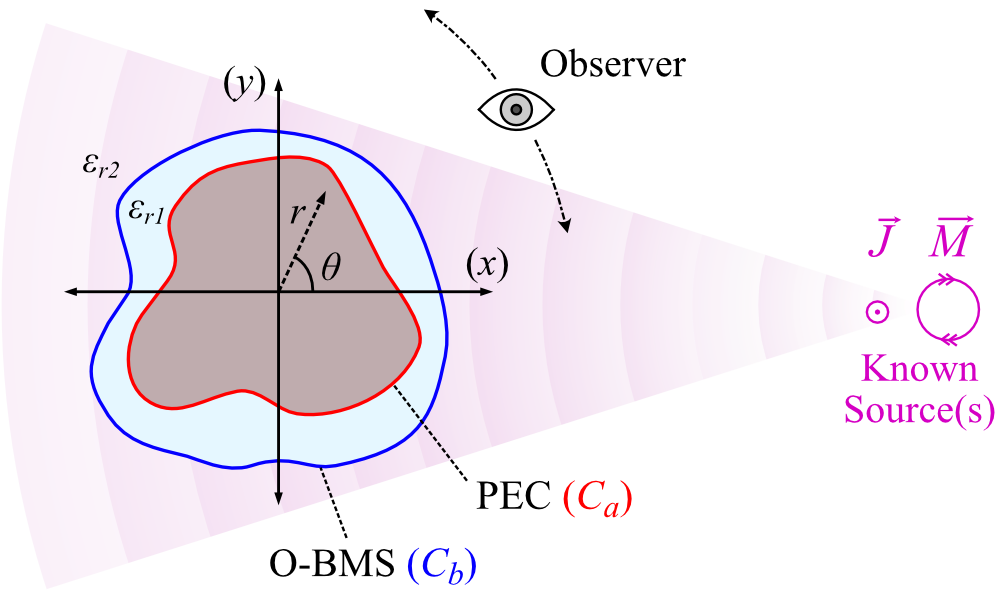}
\caption{Conformal scalar O-BMS cloak for impenetrable PEC targets.}
\label{fig:conformal_cloak_schematic}
\end{figure}

Recently, we proposed a new passive and lossless scalar metasurface configuration capable of concealing impenetrable targets~\cite{xu2020discrete}. A scalar O-BMS is placed around the target without making contact with its exterior. This permits non-zero fields on both sides of the O-BMS, allowing the prescription of an internal power profile which matches the incident power without incurring external reflections. However, the cloak as well as the target were assumed to have a circular cross sections. In this work, we expand the design space by allowing the PEC cylinder and its metasurface enclosure to take on (potentially distinct) non-circular shapes. The proposed schematic is shown in Fig.~\ref{fig:conformal_cloak_schematic}, in which the physical cross section of the PEC in the $Z$-plane is modeled by the curve $C_a$. The cross section of the metasurface is modeled by the curve $C_b$. Although it is not an inherent requirement, $C_b$ should be parallel to $C_a$ if one wishes to minimize the overall profile of the cloak. An additional advantage of parallel cross sections is that the region between the PEC and the MTS can be implemented using a flexible dielectric substrate with constant thickness. On the other hand, one may wish to shape the MTS differently from the PEC object due to structural considerations.

\begin{figure}[t]
\centering  
\subfigure[]{\includegraphics[width=0.98\linewidth]{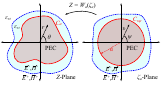}}
\\~\\
\subfigure[]{\includegraphics[width=0.98\linewidth]{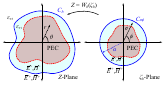}}
\caption{Conformal mappings used to enforce boundary conditions on (a) the impenetrable target; and (b) the O-BMS.}
\label{fig:PhysicalPlane_and_ComputationPlane_DoubleTransform}
\end{figure}

To engineer the electromagnetic field distributions in this problem, we must be able to simultaneously enforce the appropriate boundary conditions on $C_a$ as well as $C_b$. Since their shapes may not be related in general, we employ two independent conformal transformations, as illustrated in Fig.~\ref{fig:PhysicalPlane_and_ComputationPlane_DoubleTransform}, to handle the boundaries separately. In this figure, the $\zeta_a$-plane and the $\zeta_b$-plane are the computational planes used to enforce the boundary conditions on the PEC and the O-BMS respectively. As indicated in Fig.~\ref{fig:PhysicalPlane_and_ComputationPlane_DoubleTransform}(a), we use a complex-valued function $W_a$ to map the circle $C_{\upsilon\psi}=\{(\upsilon,\psi)|\,\upsilon=\alpha,\psi\in[0,2\pi]\}$ to $C_a$. Then, the enforcement of the condition of vanishing tangential electric field along $C_a$ is equivalent to demanding $E_z=0$ on $C_{\upsilon\psi}$. In Fig.~\ref{fig:PhysicalPlane_and_ComputationPlane_DoubleTransform}(b), we use the function $W_b$ to map the circle $C_{\sigma\phi}=\{(\sigma,\phi)|\,\sigma=\alpha,\phi\in[0,2\pi]\}$ to $C_b$. The enforcement of the BSTCs on $C_b$ then amounts to the same problem as that investigated in Sec.~\ref{sec:penetrable_targets}.

As expected, the introduction of the PEC demands new solutions to (\ref{eqn:Helmholtz}). To obtain them,  we first recognize that the problem remains unchanged from before for the external region ($\epsilon_{r2}$). Thus, the general modal solutions for the incident and the reflected fields are the same as those given in (\ref{eqn:electric_field_expressions}) and (\ref{eqn:magnetic_field_expressions}), with the new mapping $W_b$ in place of $W$. However, in the internal region ($\epsilon_{r1}$), the transmitted fields now possess different expansions owing to the PEC target. Since this region includes neither the origin nor infinity, it admits solutions in the form of linear combinations of $J_m$ and Neumann functions $Y_m$:
\begin{equation}
\label{eqn:double_map_fields}
\begin{split}
\begin{Bmatrix}E^t_z\\H^t_\parallel\end{Bmatrix} = \sum_{m=-\infty}^{\infty}&\begin{Bmatrix}A^t_{m}\\\gamma_{m}\left(W_\iota,\zeta_\iota\right)A^t_{m}\end{Bmatrix}\\
&\qquad\cdot J_m\left(k_1\left|W_\iota\left(\zeta_\iota\right)\right|\right)e^{jm\angle W_\iota\left(\zeta_\iota\right)}\\
&+\begin{Bmatrix}B^t_{m}\\\kappa_{m}\left(W_\iota,\zeta_\iota\right)B^t_{m}\end{Bmatrix}\\
&\qquad\cdot Y_m\left(k_1\left|W_\iota\left(\zeta_\iota\right)\right|\right)e^{jm\angle W_\iota\left(\zeta_\iota\right)},\\
&\iota\in\{a,b\},
\end{split}
\end{equation}
where the function compositions
\begin{equation}
\begin{split}
\gamma_{m}\left(W,\zeta\right)=&\frac{k_1}{j\omega\mu h_l}\frac{J'_m(k_1\left|W\left(\zeta\right)\right|)}{J_m(k_1\left|W\left(\zeta\right)\right|)}\cdot\frac{\partial}{\partial l}\left|W\left(\zeta\right)\right|\\
&+\frac{m}{\omega\mu h_l}\cdot\frac{\partial}{\partial l}\angle W (\zeta),\\
\kappa_{m}\left(W,\zeta\right)=&\frac{k_1}{j\omega\mu h_l}\frac{Y'_m(k_1\left|W\left(\zeta\right)\right|)}{Y_m(k_1\left|W\left(\zeta\right)\right|)}\cdot\frac{\partial}{\partial l}\left|W\left(\zeta\right)\right|\\
&+\frac{m}{\omega\mu h_l}\cdot\frac{\partial}{\partial l}\angle W (\zeta),
\end{split}
\end{equation}
are analogous to those in (\ref{eqn:modal_admittance_internal}). Under the mapping $W_{\{a,b\}}$, we have $l=\{\nu,\sigma\}$.

Using (\ref{eqn:double_map_fields}), we are able to write the electric field distributions along $C_a$ and $C_b$, respectively, in terms of vectors $\bar{E}_{z,a}$ and $\bar{E}_{z,b}$ defined as follows:

\begin{equation}
\label{eqn:double_map_E_vectors}
\begin{split}
\bar{E}_{z,\iota}^t &= \mathbf{P}^t_{A,\iota}\hat{A}^t + \mathbf{P}^t_{B,\iota}\hat{B}^t,\\
\mathbf{P}^t_{A,\iota}[n][m] &= J_{m^\star}(k_1\left|W_\iota(\zeta_n^\star)\right|)e^{jm^\star\angle W_\iota(\zeta_n^\star)},\\
\mathbf{P}^t_{B,\iota}[n][m] &= Y_{m^\star}(k_1\left|W_\iota(\zeta_n^\star)\right|)e^{jm^\star\angle W_\iota(\zeta_n^\star)},\\
 \iota&\in\{a,b\}\\
\end{split}
\end{equation}

The total tangential electric fields must vanish on $C_a$, meaning:
\begin{equation}
\label{eqn:PEC_BC}
 \hat{B}^t = -\left(\mathbf{P}^t_{B,a}\right)^{-1}\mathbf{P}^t_{A,a}\hat{A}^t.
\end{equation}
Combining (\ref{eqn:double_map_E_vectors}) and (\ref{eqn:PEC_BC}), we can rewrite the total transmitted electric field along $C_b$ as
\begin{equation}
\label{eqn:P_matrix_double_map}
\bar{E}_{z,b}^t = \left[\mathbf{P}^t_{A,b}-\mathbf{P}^t_{B,b}\left(\mathbf{P}^t_{B,a}\right)^{-1}\mathbf{P}^t_{A,a}\right]\hat{A}^t\triangleq\mathbf{P}^t\hat{A}^t.
\end{equation}
Thus, we have obtained a new definition for the matrix $\mathbf{P}^t$ specific to the current problem configuration. On the other hand, the definitions of $\mathbf{P}^{\{i,r\}}$ remain unchanged from those specified in (\ref{eqn:electric_modal_vectors}).

We also sample the transmitted magnetic fields along $C_b$, arriving at the vectorial representations as follows:
\begin{equation}
\label{eqn:double_map_H_vector}
\begin{split}
\bar{H}_{\parallel,b}^t \equiv \bar{H}_{\phi,b}^t =& \left[\mathbf{Q}^t_{A,b}-\mathbf{Q}^t_{B,b}\left(\mathbf{P}^t_{B,a}\right)^{-1}\mathbf{P}^t_{A,a}\right]\hat{A}^t\triangleq\mathbf{Q}^t\hat{A}^t,\\
\mathbf{Q}^t_{A,b}[n][m] =&\gamma_{m^\star}\left(W_b,\zeta^\star_n\right) J_{m^\star}\left(k_1\left|W_b(\zeta^\star_n)\right|\right)e^{jm^\star\angle W_b(\zeta^\star_n)}\\ 
\mathbf{Q}^t_{B,b}[n][m] =&\kappa_{m^\star}\left(W_b,\zeta^\star_n\right)Y_{m^\star}\left(k_1\left|W_b(\zeta^\star_n)\right|\right)e^{jm^\star\angle W_b(\zeta^\star_n)}\\
\end{split}
\end{equation}
Therefore, we have an updated $\mathbf{Q}^t$, whereas $\mathbf{Q}^{\{i,r\}}$ remain unchanged from (\ref{eqn:magnetic_modal_vectors}).

Having defined the new $\mathbf{P}^t$ and $\mathbf{Q}^t$, we can calculate the corresponding transmitted admittance matrix $\mathbf{Y}^t$ using (\ref{eqn:electric_magnetic_field_relations}). The matrices $\mathbf{Y}^i$ and $\mathbf{Y}^r$ can be calculated using (\ref{eqn:electric_modal_vectors}), (\ref{eqn:magnetic_modal_vectors}) and (\ref{eqn:electric_magnetic_field_relations}). However, $W_b(\zeta_n^\star)$ should be used for the function compositions $\gamma_{2,m^\star}$ and $\tau_{2,m^\star}$, since the admittance matrices are defined for fields along the curve $C_b$, which is modeled by $W_b$.

Having properly defined $\mathbf{Y}^{\{i,t,r\}}$ for the new problem, the exact same steps used in Sec.~\ref{sec:penetrable_targets} can be repeated to obtain the cloak design for impenetrable targets. To illustrate, we attempt to conceal a quasi-triangular PEC cylinder from an external plane wave traveling towards the negative $x$-direction. The cross section of the target is modeled by the mapping
\begin{equation}
\label{eqn:triangle_PEC_mapping}
Z = W_a(\zeta_a)= \frac{1}{3}\left(\zeta_a+\frac{1}{2\zeta_a^2}\right)e^{j\frac{\pi}{4}}
\end{equation}
As a simple demonstration, we choose an O-BMS cross section modeled by the mapping
\begin{equation}
\label{eqn:triangle_OBMS_mapping}
Z = W_b(\zeta_b)= \frac{1}{3}\left(\frac{10}{9}\zeta_b+\frac{1}{2\left(\frac{10}{9}\zeta_b\right)^2}\right)e^{j\frac{\pi}{4}}.
\end{equation}
The rest of the relevant parameters for this design are summarized in Tab.~\ref{tab:ex_B2_spec}. Note that our choices of $W_a$ and $W_b$ do not result in parallel $C_a$ and $C_b$, although a good approximation is achieved. Using (\ref{eqn:triangle_PEC_mapping}) and (\ref{eqn:triangle_OBMS_mapping}), the admittance matrices $\mathbf{Y}^{\{i,t,r\}}$ are populated.

\begin{table}[b]
\begin{ruledtabular}
\caption{Parameters for the quasi-triangular PEC cylinder to be concealed.}
\label{tab:ex_B2_spec}
\begin{tabular}{cccccccc}
  $f$ (GHz) & $N$ & M &$\alpha$& $\epsilon_{r1}$& $\epsilon_{r2}$\\ 
\hline
  2 & 351& 151& 1.35& 1& 1 \\ 
\end{tabular}
\end{ruledtabular}
\end{table}

\begin{figure}[t]
\centering  
\includegraphics[width=0.98\linewidth]{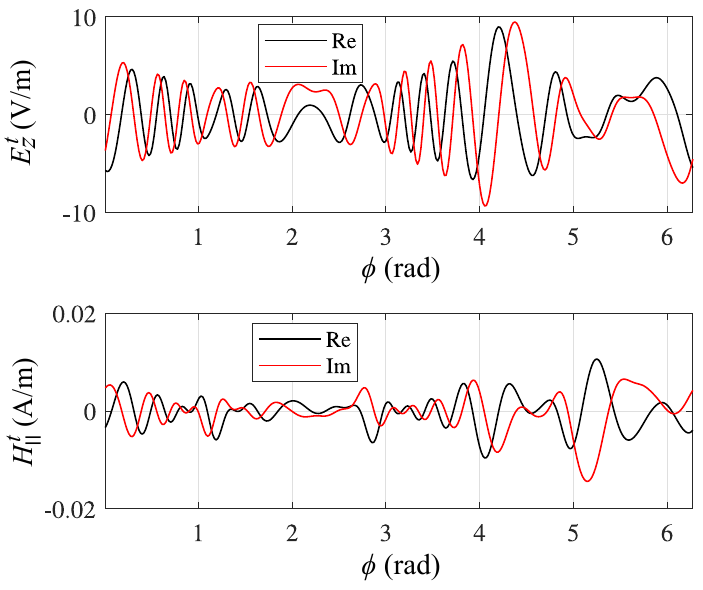}
\caption{The required transmitted auxiliary electric field $E^t_z$ and magnetic field $H^t_\parallel$($\equiv H^t_\phi$) for the passive and lossless cloaking of the quasi-triangular PEC cylinder.}
\label{fig:triangle_PEC_aux_field}
\end{figure}

Next, we solve (\ref{eqn:LPC_R}) by writing the modal expansion of the incident plane wave as~\cite{balanis}:
\begin{equation}
\hat{A}^i[m]=j^{m^\star}.
\end{equation}
Since this is a cloak, we again have $\hat{A}^r=0$. Inserting the associated fields into (\ref{eqn:LPC_R}), we solve for the required auxiliary $\hat{A}^t$, whose corresponding boundary electric and magnetic fields are plotted in Fig.~\ref{fig:triangle_PEC_aux_field}.

\begin{figure}[t]
\centering  
\includegraphics[width=0.98\linewidth]{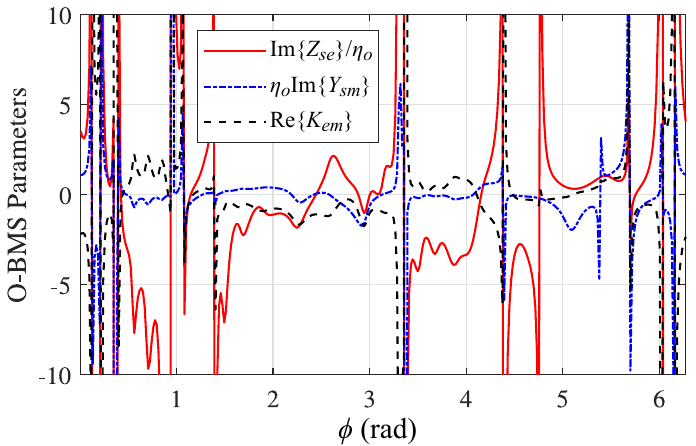}
\caption{Synthesized passive and lossless O-BMS parameters for cloaking the quasi-triangular PEC cylinder from an external plane wave.}
\label{fig:triangle_PEC_OBMS}
\end{figure}

The passive and lossless O-BMS parameters which support the complete field distributions as we have determined are plotted in Fig.~\ref{fig:triangle_PEC_OBMS}.

\begin{figure}[t]
\centering  
\subfigure[]{\includegraphics[width=0.98\linewidth]{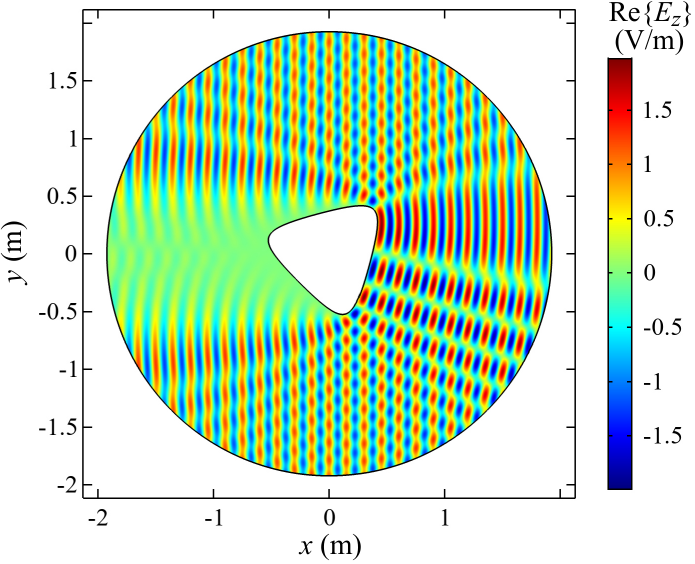}}
\\~\\
\subfigure[]{\includegraphics[width=0.98\linewidth]{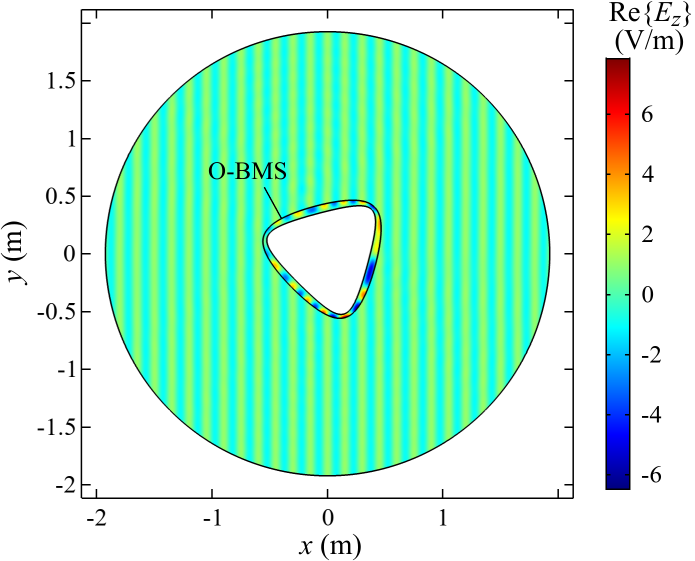}}
\caption{Snapshot of simulated $\mathrm{Re}\{E_z\}$ for a quasi-triangular PEC cylinder illuminated by a plane wave (a) without cloaking; and (b) with the O-BMS cloak.}
\label{fig:triangle_PEC_cloak_COMSOL}
\end{figure}

Implementing this design in COMSOL and simulating with an incident plane wave excitation, we obtain the total electric field distributions shown in Fig.~\ref{fig:triangle_PEC_cloak_COMSOL}. In Fig.~\ref{fig:triangle_PEC_cloak_COMSOL}(a), we show the fields without the metasurface cloak. As expected, the PEC cylinder generates a significant amount of scattering and casts a shadow due to its large electric size. In Fig.~\ref{fig:triangle_PEC_cloak_COMSOL}(b), we show the total fields with the O-BMS cloak in place. It can be seen that the external field distribution consists of just the incident plane wave and contains no observable scattering, meaning that the target has been successfully concealed. In the internal region between the PEC and the O-BMS, we see a standing wave-like field profile which serves to match the local power density of the incident plane wave in a point-wise manner.

\begin{figure}[t]
\centering  
\includegraphics[width=0.98\linewidth]{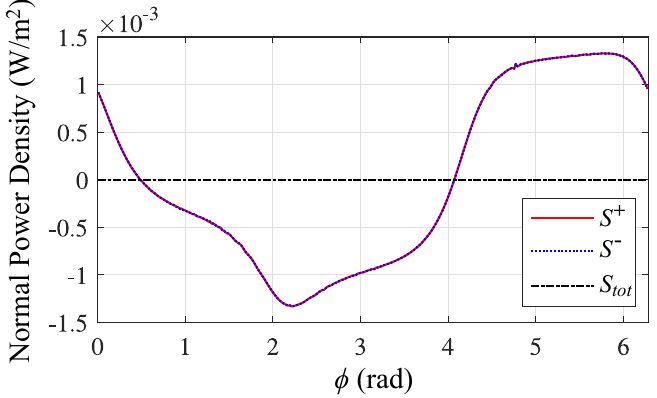}
\caption{Normal power density above and below the O-BMS cloak for the quasi-triangular PEC cylinder.}
\label{fig:triangle_PEC_powerprofile}
\end{figure}

To examine the working principles of the cloak in more detail, we plot the normal power density above ($S^+$) and below ($S^-$) the O-BMS in Fig.~\ref{fig:triangle_PEC_powerprofile}. A perfect match between the two profiles confirms that the cloak requires neither gain nor loss. Interestingly, we observe two partitions in Fig.~\ref{fig:triangle_PEC_powerprofile}: one in which the power flows outward and one in which it flows inward. This suggests that the internal auxiliary fields described by Fig.~\ref{fig:triangle_PEC_aux_field} harness the incident power on the illuminated side of the cylinder while transporting it to be emitted on the shadowed side. Therefore, the proposed cloak works on the same principle as the previously demonstrated tensorial cloaks which reroute power using surface waves~\cite{lossless_cloak}.

Lastly, we note that in this example, the thickness of the internal region has been chosen to provide a clear illustration of the working principle of the proposed design. In reality, there is no inherent restriction on the distance between the PEC and the O-BMS, which determines the overall profile of the cloak.

\subsection{Electromagnetic Illusions}
\label{sec:results_illusions}
\begin{figure}[b]
\centering  
\includegraphics[width=\linewidth]{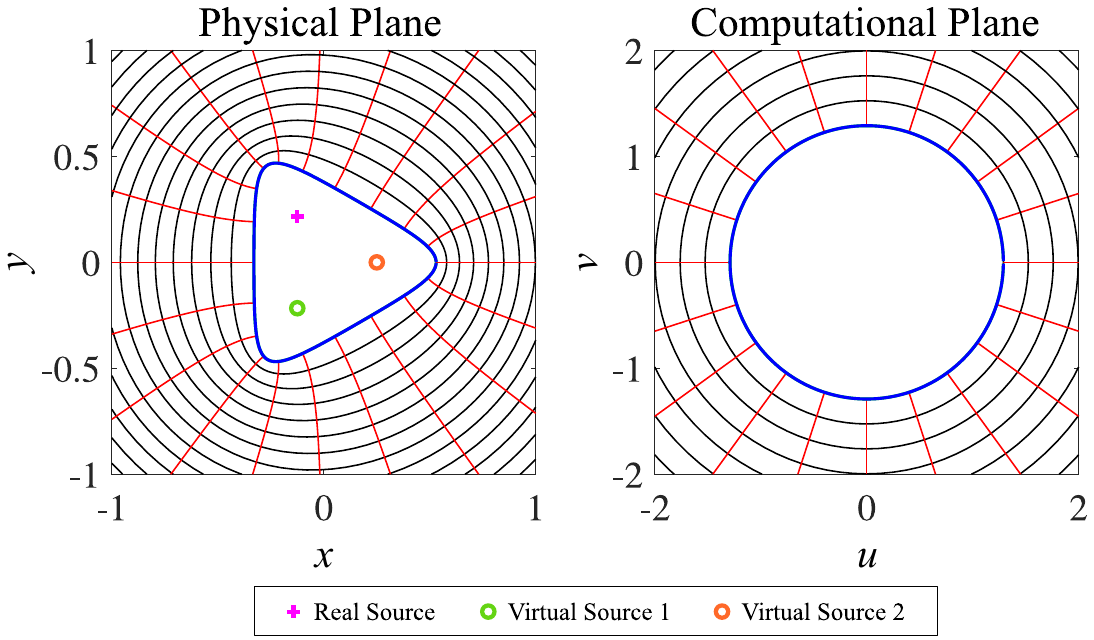}
\caption{Visualization of the conformal transformation for the quasi-triangluar illusion metasurface.}
\label{fig:triangle_illusion_conformal_visualization}
\end{figure}

As a demonstration of the capability of the proposed framework to accommodate internal sources, we design a quasi-triangular electromagnetic illusion metasurface which transforms the cylindrical wave radiated by an electric line source located at $(r,\theta)=(r_s,\theta_s)$ into the combined fields of two virtual sources located at $(r_s,\theta_s+\frac{2\pi}{3})$ and $(r_s,\theta_s+\frac{4\pi}{3})$. A schematic of the design as well as a visualization of the conformal transformation used in this problem are shown in Fig.~\ref{fig:triangle_illusion_conformal_visualization}. The geometric and material parameters for this design are summarized in Tab.~\ref{tab:ex_C_spec}.

\begin{table}[t]
\begin{ruledtabular}
\caption{Parameters for the illusion O-BMS}
\label{tab:ex_C_spec}
\begin{tabular}{ccccccccccc}
  $f$ (GHz) & $N$ & M &$\alpha$&R&q&t& $\epsilon_{r1}$& $\epsilon_{r2}$ &$r_s$(m)&$\theta_s$(rad)\\ 
\hline
 1 & 351& 71&1.29& 1/3& 2& 0 & 1& 1&0.25&$2\pi/3$\\ 
\end{tabular}
\end{ruledtabular}
\end{table}

A detailed investigation of general internally excited non-circular cylindrical MTSs can be found in Appendix~\ref{app:Internal_excitation}. Here, we simply note that the main ideas presented in the previous sections remain unchanged, and discuss some key results.

The first step of design is to populate the admittance matrices $\mathbf{Y}^{\{i,t,r\}}$ using (\ref{eqn:internal_excited_field_vectors}). Then, we need to state the known field quantities in preparation to solve the LPC equation. Unlike the previous examples, this is a transmissive metasurface, since we are judiciously prescribing the transmitted fields based on the known incident fields. 

The incident fields are those radiated by an electric line source is located at ($r,\theta$)=($r_s,\theta_s$). Since the source located in the internal region, it will be found somewhere inside the circle $C_{\sigma\phi}$. That is:
\begin{equation}
\label{eqn:line_source_int_location}
\left|W^{-1}(r_se^{j\theta_s})\right|<\alpha
\end{equation}
This suggests that we should invoke the alternate form of the addition theorem for Hankel functions~\cite{balanis}, arriving at the following modal expansion for the incident fields:
\begin{equation}
\label{eqn:line_source_int}
\hat{A}^i[m] = J_{m^\star}(k_1r_s)e^{-jm^\star\theta_s}.
\end{equation}

Since the two virtual sources also satisfy (\ref{eqn:line_source_int_location}), we can write:
\begin{equation}
\label{eqn:line_source_int2}
\hat{A}^t[m] = J_{m^\star}(k_1r_s)e^{-jm^\star\theta_s}\left\{e^{-j\frac{2\pi m^\star}{3}}+e^{-j\frac{4\pi m^\star}{3}}\right\}.
\end{equation}

\begin{figure}[t]
\centering  
\includegraphics[width=0.98\linewidth]{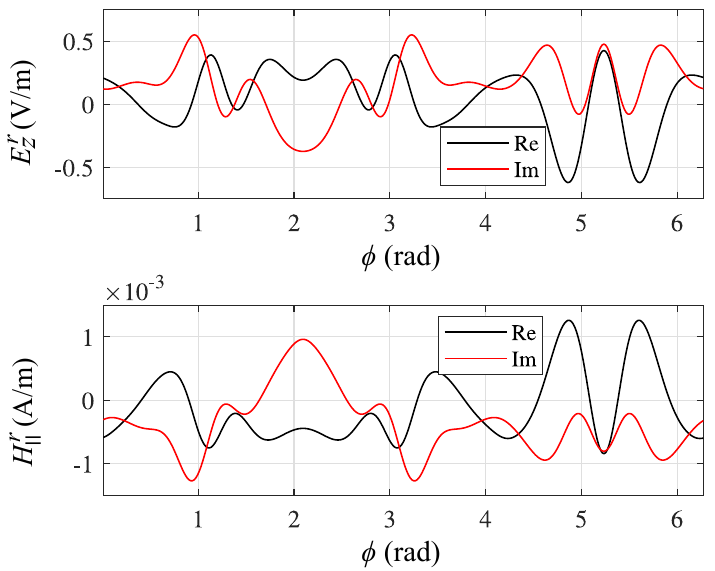}
\caption{Required auxiliary reflection for the quasi-triangular illusion metasurface shown in Fig.~\ref{fig:triangle_illusion_conformal_visualization}.}
\label{fig:triangle_aux_field}
\end{figure}

To guarantee passivity and losslessness of the design, we insert the fields associated with (\ref{eqn:line_source_int}) and (\ref{eqn:line_source_int2}) into the transmissive LPC equation (\ref{eqn:LPC_T}). We solve for the required $\hat{A}^r$ and plot its associated boundary electromagnetic fields in Fig.~\ref{fig:triangle_aux_field}.

\begin{figure}[t]
\centering  
\includegraphics[width=0.98\linewidth]{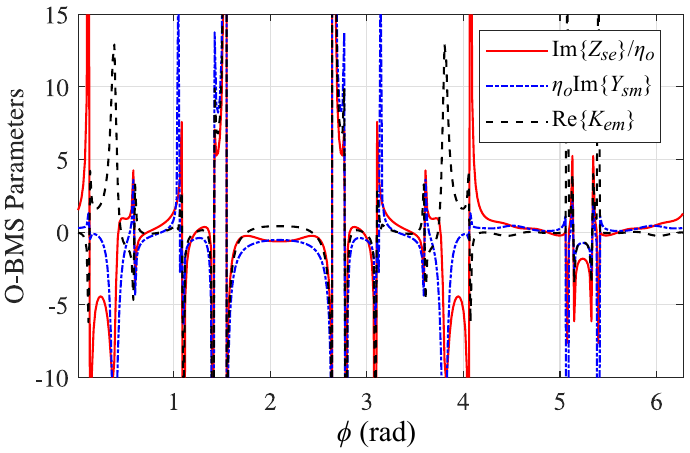}
\caption{Synthesized passive and lossless O-BMS parameters for the quasi-triangular illusion metasurface.}
\label{fig:triangle_OBMS}
\end{figure}

To synthesize the required O-BMS, (\ref{eqn:synthesis}) is used in conjunction with the updated field definitions of $\bar{E}^\pm_z$ and $\bar{H}^\pm_\phi$ provided in (\ref{eqn:fields_int_excited}). The solved passive and lossless surface parameters are plotted in Fig.~\ref{fig:triangle_OBMS}.

\begin{figure}[ht!]
\centering  
\subfigure[]{\includegraphics[width=0.85\linewidth]{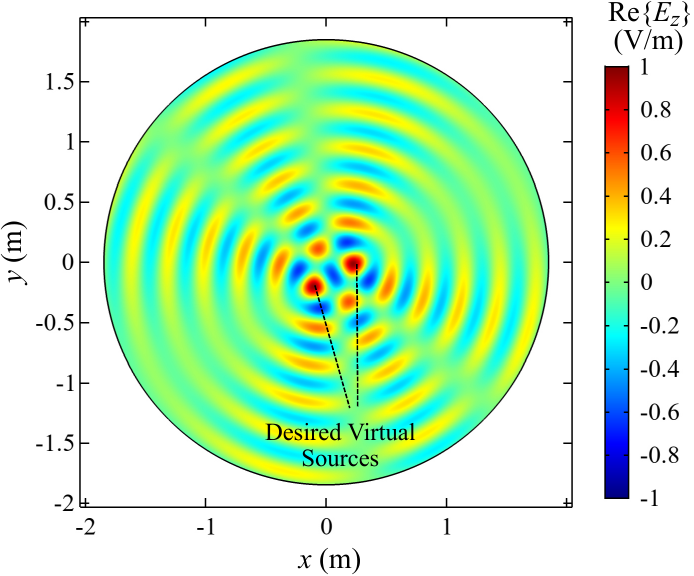}}
\\~\\
\subfigure[]{\includegraphics[width=0.85\linewidth]{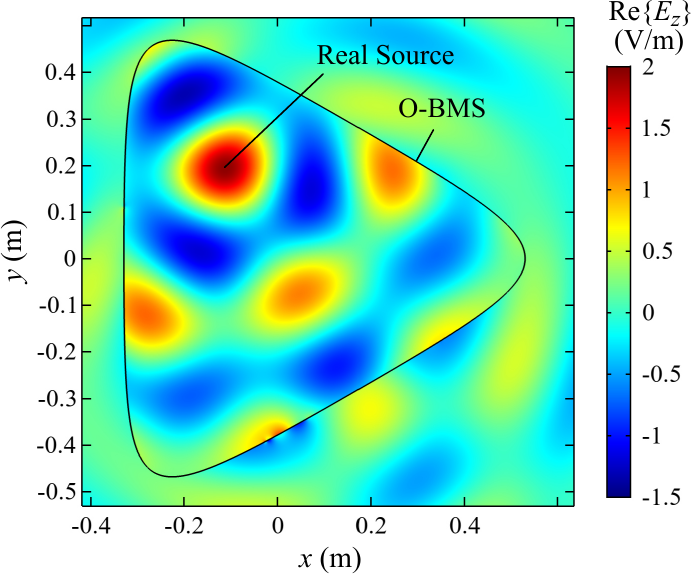}}
\\~\\
\subfigure[]{\includegraphics[width=0.85\linewidth]{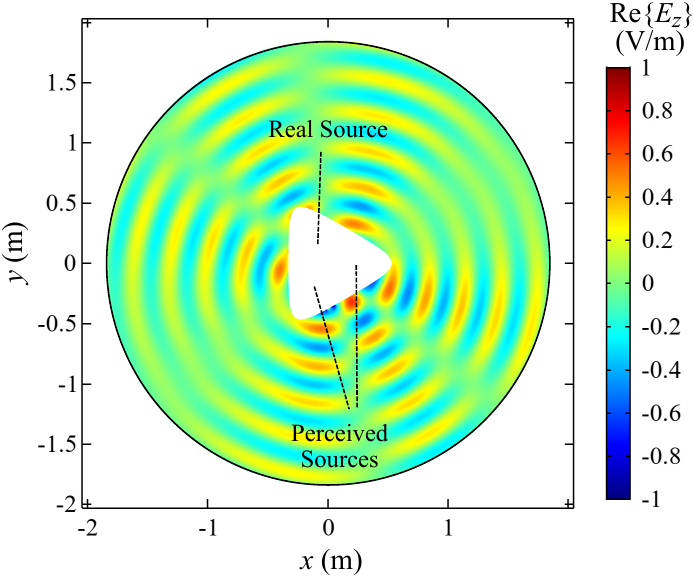}}
\caption{(a) Desired field distribution of two line sources. Snapshot of imulated $\mathrm{Re}\{E_z\}$ (b) inside and (c) outside of the MTS cavity.}
\label{fig:triangle_illusion_COMSOL}
\end{figure}

We implement the O-BMS in COMSOL and excite the real line source. Snapshots of the resulting electric field distribution are shown in Fig.~\ref{fig:triangle_illusion_COMSOL}. The internal (Fig.~\ref{fig:triangle_illusion_COMSOL}(b)) and external (Fig.~\ref{fig:triangle_illusion_COMSOL}(c)) fields are plotted separately, to provide a better comparison with the desired field distribution radiated by the two virtual sources (Fig.~\ref{fig:triangle_illusion_COMSOL}(a)). We observe an almost exact match for the external fields in Fig.~\ref{fig:triangle_illusion_COMSOL}(a) and Fig.~\ref{fig:triangle_illusion_COMSOL}(c), which suggests that an external observer would perceive two displaced virtual sources, instead of the single real source.

\section{Conclusions}
We presented methods for analyzing and designing cylindrical metasurfaces with non-circular cross sections based on conformal transformations. The physical space inhabited by the irregularly shaped metasurface is mapped to a new computational space in which the surface resides on a perfectly circular cylinder. In the computational space, it is straightforward to enforce the bianisotropic sheet transition conditions while solving the Helmholtz wave equation, owing to the canonical shape of the boundary. This admits closed-form analysis and synthesis equations for various cylindrical metasurface-based devices with unique cross-sectional shapes. Importantly, since the problem is solved analytically using modal expansions, we can efficiently identify field distributions that satisfy local power conservation, guaranteeing passivity and losslessness of the generated designs. 

Finite element simulations were conducted to confirm the accuracy of the proposed analysis method. Several passive and lossless devices with different practical applications including cloaking and electromagnetic illusions were demonstrated with the help of the proposed design procedure. Their functionalities are verified using finite element simulations.

\vfill
\bibliography{Conformal_MTS_V2}
\newpage
\appendix
\section{Internally Excited Conformal O-BMS}
\label{app:Internal_excitation}
\begin{figure}[h]
\centering
\includegraphics[width=0.9\linewidth]{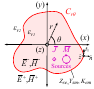}
\caption{Internally excited cylindrical O-BMS with non-circular cross section.}
\label{fig:PhysicalPlane2}
\end{figure}

In this appendix, we extend the framework presented in the main text, in order to model internal sources. A general internally excited cylindrical scalar metasurface is depicted in Fig.~\ref{fig:PhysicalPlane2}.

Due to the new source location, the incident and the reflects fields are now found inside $C_{r\theta}$ while the transmitted fields are outside. In other words, in the computational plane, we have
\begin{equation}
\begin{split}
\label{eqn:fields_int_excited}
\bar{E}^+_z = \bar{E}^t_z,\quad \bar{E}^-_z = \bar{E}^i_z + \bar{E}^r_z,\\
\bar{H}^+_\phi = \bar{H}^t_\phi,\quad \bar{H}^-_\phi = \bar{H}^i_\phi + \bar{H}^r_\phi.
\end{split}
\end{equation}
In the following sections, we explore the implications of (\ref{eqn:fields_int_excited}) with respect to the analysis and the synthesis equations.

\subsection{Analysis}
\label{app:analysis}
A consequence of the interchanged locations of the incident, transmitted and reflected fields is that the general solution to (\ref{eqn:conformal_Maxwells_eqn}) is now given by the modal expansions
\begin{equation}
\label{eqn:internal_excited_field_expressions}
\begin{split}
E^i_z &= \sum_{m=-\infty}^{\infty}A^i_mH^{(2)}_m(k_1\left|W(\zeta)\right|)e^{jm\angle W(\zeta)},\\
E^t_z &= \sum_{m=-\infty}^{\infty}A^t_mH^{(2)}_m(k_2\left|W(\zeta)\right|)e^{jm\angle W(\zeta)},\\
E^r_z &= \sum_{m=-\infty}^{\infty}A^r_mJ_m(k_1\left|W(\zeta)\right|)e^{jm\angle W(\zeta)},\\
H^i_\phi &= \sum_{m=-\infty}^{\infty}\tau_{1,m}\left(W,\zeta\right)A^i_mH^{(2)}_m(k_1\left|W\left(\zeta\right)\right|)e^{jm\angle W (\zeta)},\\
H^t_\phi &= \sum_{m=-\infty}^{\infty}\tau_{2,m}\left(W,\zeta\right)A^t_mH^{(2)}_m(k_2\left|W\left(\zeta\right)\right|)e^{jm\angle W (\zeta)},\\
H^r_\phi &= \sum_{m=-\infty}^{\infty}\gamma_{1,m}\left(W,\zeta\right)A^r_mJ_m(k_1\left|W\left(\zeta\right)\right|)e^{jm\angle W (\zeta)},\\
\end{split}
\end{equation}
which can be sampled at $N$ equally spaced points along the circle $\left|\zeta\right|=\alpha$ in the computational plane, and truncated to $M$ modes, forming the vectors
\begin{equation}
\label{eqn:internal_excited_field_vectors}
\begin{split}
\bar{E}^{\{i,t,r\}}_z =& \mathbf{P}^{\{i,t,r\}}\hat{A}^{\{i,t,r\}},\quad\bar{H}^{\{i,t,r\}}_\phi = \mathbf{Q}^{\{i,t,r\}}\hat{A}^{\{i,t,r\}}\\
\mathbf{P}^i[n][m] =&H^{(2)}_{m^\star}(k_1\left|W(\zeta^\star_n)\right|)e^{jm^\star\angle W(\zeta^\star_n)},\\
\mathbf{P}^t[n][m] =&H^{(2)}_{m^\star}(k_2\left|W(\zeta^\star_n)\right|)e^{jm^\star\angle W(\zeta^\star_n)},\\
\mathbf{P}^r[n][m] =&J_{m^\star}(k_1\left|W(\zeta^\star_n)\right|)e^{jm^\star\angle W(\zeta^\star_n)},\\
\mathbf{Q}^i[n][m] =&\tau_{1,m^\star}\left(W,\zeta^\star_n\right)H^{(2)}_{m^\star}\left(k_1\left|W(\zeta^\star_n)\right|\right)e^{jm^\star\angle W(\zeta^\star_n)},\\
\mathbf{Q}^t[n][m] =&\tau_{2,m^\star}\left(W,\zeta^\star_n\right)H^{(2)}_{m^\star}\left(k_2\left|W(\zeta^\star_n)\right|\right)e^{jm^\star\angle W(\zeta^\star_n)},\\
\mathbf{Q}^r[n][m] =&\gamma_{1,m^\star}\left(W,\zeta^\star_n\right)J_{m^\star}\left(k_1\left|W(\zeta^\star_n)\right|\right)e^{jm^\star\angle W(\zeta^\star_n)}.
\end{split}
\end{equation}
Then, the boundary electric and magnetic field vectors can be related to each other algebraically with the help of the admittance matrices  $\mathbf{Y}^{\{i,t,r\}}$ defined in (\ref{eqn:electric_magnetic_field_relations}).

In addition to the new expressions for $\mathbf{Y}^{\{i,t,r\}}$, the discretized BSTCs (\ref{eqn:BSTC_discrete_analysis}) must be updated to reflect the transposed field locations specified by (\ref{eqn:fields_int_excited}). This leads to new formulas for the modal transmission and reflection matrices $\mathbf{T}$ and $\mathbf{R}$ as follows:

\begin{equation}
\label{eqn:TR_int}
\begin{split}
\mathbf{T} =& \left(\mathbf{P}^t\right)^{-1}\mathbf{t}_a^{-1}\mathbf{t}_b\mathbf{P}^i,\\
\mathbf{R} =& \left(\mathbf{P}^r\right)^{-1}\mathbf{r}_a^{-1}\mathbf{r}_b\mathbf{P}^i,\\
\mathbf{t}_a=&\left(\frac{1}{2}\mathbf{I}+\mathbf{Z}\mathbf{Y}^r-\mathbf{K}\right)^{-1}\left(\frac{1}{2}\mathbf{I}-\mathbf{Z}\mathbf{Y}^t+\mathbf{K}\right)\\
&-\left(\frac{1}{2}\mathbf{Y}^r+\mathbf{Y}+\mathbf{K}\mathbf{Y}^r\right)^{-1}\left(\frac{1}{2}\mathbf{Y}^t-\mathbf{Y}-\mathbf{K}\mathbf{Y}^t\right),\\
\mathbf{t}_b=&\left(\frac{1}{2}\mathbf{Y}^r+\mathbf{Y}+\mathbf{K}\mathbf{Y}^r\right)^{-1}\left(\frac{1}{2}\mathbf{Y}^i+\mathbf{Y}+\mathbf{K}\mathbf{Y}^i\right)\\
&-\left(\frac{1}{2}\mathbf{I}+\mathbf{Z}\mathbf{Y}^r-\mathbf{K}\right)^{-1}\left(\frac{1}{2}\mathbf{I}+\mathbf{Z}\mathbf{Y}^i-\mathbf{K}\right),\\
\mathbf{r}_a=&\left(\frac{1}{2}\mathbf{I}-\mathbf{Z}\mathbf{Y}^t+\mathbf{K}\right)^{-1}\left(\frac{1}{2}\mathbf{I}+\mathbf{Z}\mathbf{Y}^r-\mathbf{K}\right)\\
      &-\left(\frac{1}{2}\mathbf{Y}^t-\mathbf{Y}-\mathbf{K}\mathbf{Y}^t\right)^{-1}\left(\frac{1}{2}\mathbf{Y}^r+\mathbf{Y}+\mathbf{K}\mathbf{Y}^r\right),\\
\mathbf{r}_b=&\left(\frac{1}{2}\mathbf{Y}^t-\mathbf{Y}-\mathbf{K}\mathbf{Y}^t\right)^{-1}\left(\frac{1}{2}\mathbf{Y}^i+\mathbf{Y}+\mathbf{K}\mathbf{Y}^i\right)\\
	&-\left(\frac{1}{2}\mathbf{I}-\mathbf{Z}\mathbf{Y}^t+\mathbf{K}\right)^{-1}\left(\frac{1}{2}\mathbf{I}+\mathbf{Z}\mathbf{Y}^i-\mathbf{K}\right).\\
\end{split}
\end{equation}
\vfill

\subsection{Synthesis}
To design an internally excited O-BMS, we first note that the LPC equation (\ref{eqn:LPC}) assumes nothing about the source location. Therefore, to ensure passivity and losslessness, we can reuse (\ref{eqn:LPC_T}) or (\ref{eqn:LPC_R}), as long as we incorporate the appropriate problem-specific definitions of $\{\mathbf{P}^r,\mathbf{Q}^r\}$ or $\{\mathbf{P}^t,\mathbf{Q}^t\}$ as specified in (\ref{eqn:internal_excited_field_vectors}).

Once the complete electromagnetic fields at the O-BMS interface have been determined, the synthesis equations (\ref{eqn:synthesis}) can be used in conjunction with (\ref{eqn:fields_int_excited}).

\end{document}